\documentstyle[preprint,prb,epsf,aps]{revtex}
\tighten
\begin{document}
\title{Ultrafast Coulomb-induced dynamics of 2D magnetoexcitons}
\author{T. V. Shahbazyan and N. Primozich}
\address{Department of Physics and Astronomy,  
Vanderbilt University, Nashville, TN 37235}
\author{I. E. Perakis}
\address{Department of Physics and Astronomy,  
Vanderbilt University, Nashville, TN 37235
and Department of Physics, University of Crete, 
P.O. Box 2208, 710 03, Heraklion, Crete, Greece}
\maketitle
\begin{abstract}

We study theoretically the ultrafast nonlinear optical 
response of quantum well excitons in a perpendicular magnetic
field. We show that for magnetoexcitons confined to the lowest
Landau levels, the third-order four-wave-mixing (FWM) 
polarization is dominated by the exciton-exciton interaction 
effects. For repulsive interactions, we identify two regimes 
in the time-evolution of the  optical polarization characterized  
by exponential and {\em power law} decay of the FWM signal.
We describe these regimes by deriving an analytical solution for 
the memory kernel of the two-exciton wave-function in strong 
magnetic field. For strong exciton-exciton interactions, the 
decay of the FWM signal is governed by an antibound resonance 
with an interaction-dependent decay rate. For weak interactions, 
the continuum of exciton-exciton scattering states leads to a 
long tail of the time-integrated FWM signal for negative time 
delays, which is described by the product of a power law 
and a logarithmic factor. By combining this analytic solution 
with numerical calculations, we study the crossover between 
the exponential and non-exponential regimes as a function of 
magnetic field. For attractive exciton-exciton interaction, we 
show that the time-evolution of the FWM signal is dominated by 
the biexcitonic effects.  
\end{abstract}


\section{introduction}
\label{sec:introduction}

During recent years, the role of many-body correlations has been 
a dominant theme in the studies of the ultrafast nonlinear optical
response of semiconductors in the coherent regime.\cite{chemla99}
In the lowest order in the optical fields, the nonlinear optical
polarization comes from  the excitation of two electron-hole 
({\em e-h}) pair states. 
The Coulomb interactions between these elementary excitations govern
the spectral and temporal properties of the
nonlinear optical response.\cite{haugkoch,mukamel-book,shah96}

In noninteracting systems, such as atom-like two-level  systems, 
the only source of optical nonlinearity is the Pauli blocking. 
This introduces local correlations, due to the phase space filling, 
between the carriers excited by the pump and the probe optical
pulses, which are separated by a time delay $\tau$.\cite{yajima79}
The Coulomb interaction brings about new nonlinearities
by changing the dynamics of the photoexcited {\em e-h} pairs. 
The simplest theoretical approach that accounts for the interactions
between photoexcited {\em e-h} pairs
is the time-dependent Hartree-Fock approximation, which leads to the
well known semiconductor Bloch equations (SBE).
\cite{s-rink86,s-rink88,lindberg88,lindberg89,binder91,haugkoch}
The latter have been successfully used to describe a variety of
experimental results, including, e.g., the ac Stark effect 
and Rabi oscillations,\cite{s-rink88,binder91,haugkoch} 
the fast exponential decay of the time-integrated (TI) 
four-wave-mixing (FWM) signal for negative time delays
($\tau<0$),\cite{leo90,wegener90} the time-dependence of the
time-resolved (TR)  FWM signal\cite{leo90,wegener90,weiss92}, the
photon echoes coming from the continuum of states,\cite{lindberg92,glutch95}, 
and some aspects of phase measurements.\cite{bigot93,chemla94}
Being a mean-field approach, the SBE's replace the Coulomb interactions
by a  spatially-averaged effective field of all the photoexcited
excitons, and thus neglect correlations due
to exciton-exciton scattering and bound biexciton states.

More recent experiments indicated the importance of
Coulomb  correlations, e.g., in the dependence of the nonlinear
response on the polarization of the optical fields.\cite{axt98,chemla99} 
Many of the experimental results have been explained in terms
of excitation-induced dephasing,\cite{wang93,wang94,hu94,rappen94}
and exciton-exciton scattering
processes.\cite{lindberg94,axt95,schafer96} The importance of  
bound biexciton states in FWM was also pointed out, e.g., in 
Refs.\onlinecite{feuerbacher91,bar-ad92,lovering92,pantke93,bott93,mayer94,mayer95}.
Exciton-exciton correlations have been shown to play a significant
role in the ac Stark effect\cite{sieh99} and in quantum beats.\cite{aoki99}
Several systematic methods for incorporating many-particle
correlations in the coherent nonlinear optical
response have been developed,\cite{axt98}
including the effective exciton Hamiltonian
approach,\cite{spano91,leegwater92,chernyak93,mukamel-book} 
the coherently controlled truncation
scheme,\cite{balslev89,axt94,victor95,schafer96}, 
the correlation function approach,\cite{ost95,ost98,ost99} and
the canonical transformation approach.
\cite{per94,perakis99,primozich00,shahbazyan00,ssr,per96,per-chem}

In this paper, we study the role of exciton-exciton 
correlations on the ultrafast nonlinear optical response 
of quantum well (QW) excitons in a perpendicular magnetic
field ${\bf B}$. The effect of a magnetic field on the 
photoexcited carriers is twofold. 
First, the conduction and valence bands collapse into
discrete sets of highly  degenerate Landau levels (LL). 
The magnetic field squeezes the electron and hole wave functions
(magnetic confinement).
The relative importance of the Coulomb versus 
magnetic confinement in determining the exciton states 
depends on the ratio  $\omega_{c}/2 E_{B}=(a_{B}/l)^{2}$,  
where $a_{B}$ and $ E_{B}$ are 
the exciton Bohr radius and binding energy at zero field,
while  $l=\sqrt{1/eB}$ and $\omega_{c}$ are the magnetic length
and the cyclotron frequency, respectively (we set $\hbar=1$).
With increasing magnetic field, the crossover from Coulomb-bound
excitons to magnetoexcitons occurs for $l\lesssim a_B$, where
the binding energy of the 2D magnetoexciton, $E_0\sim e^2/l$,
exceeds the Bohr energy $E_B \sim e^{2}/a_{B}$.
In typical GaAs/AlGaAs quantum wells, this crossover
to the  strong magnetic field regime occurs for $B\gtrsim 4$ T.

The second effect of the magnetic field is that, by freezing 
the kinetic energy of the photoexcited carriers in two spatial
dimensions, it increases the relative strength of the exciton-exciton
interactions.\cite{stafford90,glutsch95,stark90,carmel93,rappen91,cundiff96,jiang93,siegner94,siegner95,kner97,kner98,kner99,chernyak98,yoko99}
In bulk semiconductors, FWM experiments showed a
strong enhancement of the FWM signal with increasing 
magnetic field.\cite{kner97,kner98,kner99}
At the same time, the time-integrated signal exhibited 
a long non-Markovian decay for negative time
delays.\cite{kner97,kner98,kner99}

In QW's, the exciton-exciton interactions are expected to dominate the
coherent optical dynamics as the magnetic field increases. 
When the LL separation exceeds the
magnetoexciton binding energy, $\omega_c/|E_0|\sim a_B/l \gg 1$, the
inter-LL transitions are suppressed and the optical response is
determined by magnetoexcitons confined to the lowest LL (LLL).
Note that, in an ideal 2D system, the interactions
between magnetoexcitons with zero center of mass 
momentum are suppressed due to the electron-hole
symmetry.\cite{lerner81,bychkov83,paquet85} In  
realistic QW's, however, this symmetry is lifted due to, e.g., the
differing band offsets, lateral confinement, and disorder.\cite{qhe}
As was first pointed out in Refs.\onlinecite{chernyak98,yoko99}, 
the {\em e-h} asymmetry plays an important role in the nonlinear
response of magnetoexcitons in QW's.

We investigate here the role of four-particle correlations in the
coherent ultrafast dynamics of 2D magnetoexcitons confined to the LLL. 
Using the canonical transformation formalism
\cite{perakis99,primozich00,ssr},
we show that, due to the discrete nature of LL's, the third-order 
FWM polarization is determined by a momentum-independent two-exciton
amplitude $\chi(t)$, satisfying a simple equation
\begin{eqnarray} 
\label{eq-chi1}
i\partial_t \chi(t)-2\Omega_0\chi(t)=
V^{HF}p^2(t) + F(t),
\end{eqnarray}
where $\Omega_0$ is the pump detuning from the magnetoexciton 
level. The first source term in the rhs represent the two pump-induced
polarizations, $p(t)$, interacting via the Hartree-Fock potential
$V^{HF}$, while $F(t)$ originates from the exciton-exciton correlations. 

We have calculated $F(t)$ numerically and showed that, for a wide
range of parameters, the results can be described in terms of  an 
analytical solution derived for strong magnetic field. 
In particular, $\chi(t)$ can be expressed in terms of 
the Hartree-Fock solution $\chi^{HF}(t)$ [obtained by setting
$F(t)=0$ in Eq.\ (\ref{eq-chi1})] as follows: 
\begin{equation}
\label{chi-final1}
\chi(t)=\int_{-\infty}^{\infty}dt'S(t-t')\chi^{HF}(t'),
\end{equation}
where $S(t)$ represents a  {\em memory kernel} whose time dependence is
governed  by the exciton-exciton correlations.
For  sufficiently strong magnetic fields, such that the  magnetic
length $l$, is smaller than
the chacteristic range of the exciton-exciton potential, $a$, 
we have derived a simple analytic expression for
the memory function $S(\omega)$ in the frequency domain,
that incorporates both the biexcitonic and the 
exciton-exciton scattering effects nonperturbatively.  

For attractive exciton-exciton interaction, 
the biexciton effects dominate.
For $l \ll a$, when the momentum exchange between excitons is
suppressed, we recover the average polarization model 
(APM).\cite{wegener90,ssr91,schafer96,chemla99,kner99}
For weaker fields, we show that the time-evolution of the FWM
signal is determined by both biexcitonic and exciton-exciton
scattering effects.
For short pump durations, we reproduce the 
biexciton oscillations in the TI-FWM signal.\cite{yoko99}
As the pump duration becomes longer than 
the biexciton binding energy, we find a crossover between two
exponential regimes in the SR- and TI-FWM spectra, originating from
different dephasing rates of the exciton and biexciton bound states.

For repulsive exciton-exciton interactions, we find two
regimes in the time dependence of the FWM signal governed by a
dimensionless parameter $G\sim u_0a^2/E_0l^2$, where $E_0$ is
the magnetoexciton binding energy and $u_0$ is the
characteristic strength of the exciton-exciton interactions ($u_0<0$
for repulsive interactions). 
For $|G|\gg 1$, the memory function is dominated by an antibound
resonance, made up from the continuum of the exciton-exciton states,
with the width $\gamma_{sc}\sim u_0Ge^{-|G|}$. In 
this regime, the TI-FWM signal exhibits oscillations with a period
determined by the energy of an antibound state, together with an
overall exponential decay characterized by the {\em interaction-induced}
time $\gamma_{sc}^{-1}$. For $|G|\ll 1$, the long-time behavior of the
FWM signal is determined by the singular (logarithmic) low-frequency
behavior of $S(\omega)$; this leads to a {\em power low} decay of the
TI-FWM signal for negative time delays: TI-FWM$\propto \tau^{-2}$ (up
to log factor). In the intermediate case $|G|\sim 1$, the crossover
between the two regime as a function of time delay occurs on the scale
$|\tau|\sim |u_0|^{-1}$. 
By combining our analytic solution with numerical
calculations, we study the crossover between exponential and
non-exponential regimes as a function of magnetic field. 
Our results demonstrate that four-particle Coulomb correlations 
lead to a qualitatively different dynamics of the two-exciton
excitations as compared to the one-exciton states.

The paper is organized as follows.  
In Section \ref{sec:magnex}, we  summarize relevant facts regarding  2D
excitons in 
a strong magnetic field. In Section \ref{sec:fwm} we present an expression 
for the FWM polarization obtained in 
Appendices \ref{app:gen} and \ref{app:fwm}.
In Section \ref{sec:analytic}, we derive an analytical solution for the
two-exciton amplitude in the case of strong 
magnetic field and in Section \ref{sec:memory} we analyze 
the memory function in all the regimes of interest.
In Section \ref{sec:numeric}, we presents the results of numerical 
calculations for the FWM polarization and compare them with our
analytical solution.  
Section \ref{sec:conclusion} concludes the paper. 

\section{2D excitons in strong magnetic field}
\label{sec:magnex}

In this section, we summarize some basic facts regarding 2D
magnetoexcitons.\cite{lerner81,bychkov83,paquet85} 
We consider the case when the magnetic field is sufficiently strong
and the central optical frequency is tuned to transitions between
the lowest Landau levels of the conduction and valence bands. In this case,
the inter-LL transitions are suppressed so that
the photoexcited {\em e-h} pairs are restricted to the LLL. 
In the Landau gauge, ${\bf A}=(0,Bx)$,
the electron and hole wave-functions, $\psi_k$  and $\bar{\psi}_k$, are
characterized by the  $y$-component of the momentum, $k=-x_0/l^2$, 
where $x_0$ is the $x$-coordinate of the cyclotron orbit center, 
\begin{equation}
\psi_k({\bf r})=\frac{1}{\sqrt{L\pi^{1/2}l}}
\, e^{iky-(x+kl^2)^2/2l^2},
~~~
\bar{\psi}_k({\bf r})=\psi_{-k}^{\ast}({\bf r}),
\end{equation}
($L$ is the system size).
The normalized  one-exciton wave function, 
with total momentum ${\bf p}$, 
can be  expressed in terms of the electron and hole wave functions 
as\cite{lerner81,bychkov83,paquet85} 
\begin{equation}
\label{Psi-exc}
\Psi_{\bf p}({\bf r}_1, {\bf r}_2)
=N^{-1/2}\sum_ke^{-ikp_xl^2}\psi_{p_y/2+k}({\bf r}_1)
\bar{\psi}_{p_y/2-k}({\bf r}_2)
=\frac{N^{-1/2}}{2\pi l^2}
e^{i{\bf p\cdot R}-iXy/l^2-({\bf r}+l^2{\bf p\times z})^2/4l^2},
\end{equation}
where ${\bf r}={\bf r}_1-{\bf r}_2$, 
${\bf R}=({\bf r}_1+{\bf r}_2)/2$ are the relative and  the 
average coordinates, respectively, $N=L^2/2\pi l^2$ is the LL
degeneracy, and ${\bf z}$ is the unit vector in
the direction of the magnetic field. 
The magnetoexciton creation operator
is given in terms of the electron and hole operators 
$a_k^{\dag}$ and $b_k^{\dag}$, 
by the  similar expression,
\begin{equation}
\label{d}
d_{\bf p}^{\dag}
=N^{-1/2}\sum_ke^{-ikp_xl^2}a_{p_y/2+k}^{\dag}b_{p_y/2-k}^{\dag}.
\end{equation}

Consider first the spectrum of a single magnetoexciton.
In the one-exciton basis, the Hamiltonian 
$H_{eh}^{(1)}$ that describes 
the interaction between an electron and a hole, 
is diagonal: 
\begin{equation}
\label{H-exc}
H_{eh}^{(1)}
=\sum_{\bf p}\varepsilon_p \, d_{\bf p}^{\dag}d_{\bf p},
\end{equation}
where $\varepsilon_p$ is the 2D magnetoexciton energy, given by
\begin{equation}
\label{spectr}
\varepsilon_p=
-\int \frac{d{\bf q}}{(2\pi)^2}v_qe^{-q^2l^2/2+i({\bf q}\times{\bf p})
\cdot {\bf z}l^2}
=-\frac{e^2}{\kappa l}\sqrt{\frac{\pi}{2}}e^{-p^2l^2/4}I_0(p^2l^2/4).
\end{equation}
Here $v_q=2\pi e^2/\kappa q$ is the 2D Coulomb potential ($\kappa$ is
the dielectric constant) and $I_0(x)$ 
is the Bessel function.
For small center of mass momenta, $pl\ll 1$, the magnetoexciton 
spectrum is parabolic,
\begin{equation}
\label{spectrum}
\varepsilon_p=-E_0+\frac{p^2}{2M},
~~~
M=\frac{8}{3}\sqrt{\frac{2}{\pi}}\frac{\kappa}{e^2 l},
\end{equation}
where $E_0=\frac{e^2}{\kappa l}\sqrt{\frac{\pi}{2}}$  is 
the magnetoexciton binding energy and $M$ is the magnetoexciton 
effective mass. For large momenta, $pl\gg 1$, 
Eq.\ (\ref{spectr}) yields $\varepsilon_p=-e^2/\kappa pl^2$. Since for 
strong magnetic field the exciton momentum is proportional to
the {\em e-h} separation, $p\sim r/l^2$, for large momenta we recover
the Coulomb potential between a well-separated electron and hole,
$-e^2/\kappa r$. On the other hand, for small momenta, 
the electron and hole sit on top of each other forming 
a neutral magnetoexciton with the effective mass 
$M$.\cite{lerner81,bychkov83,paquet85}
The one-exciton Hamiltonian is 
$H_X=H_0+H_{eh}^{(1)}$, where $H_{0}$ 
describes a noninteracting {\em e-h} pair with energy
$E_0^e+E_0^h+E_g-\omega_0=\bar{E}_g-\omega_0$, where 
$E_0^e$ and $E_0^h$ are the electron and hole LLL energies,
respectively, $E_g$ is the semiconductor bandgap, and $\omega_0$ is
the pump central frequency (we are working 
in the rotating frame).
The eigenstates of $H_{X}$,  $|{\bf p}\rangle= d_{\bf p}^{\dag}|0\rangle$, 
satisfy the eigenvalue equation
\begin{equation}
\label{HX}
H_X|{\bf p}\rangle=
\Omega_p|{\bf p}\rangle,
~~~
\Omega_p=\bar{E}_g + \varepsilon_p-\omega_0.
\end{equation}
Note that the {\em linear} response in strong field 
is similar to that of a two-level system. 
This can be seen, e.g., by noticing that the optical transition
operator, $U^{\dag}$ is proportional to the exciton creation operator
[setting ${\bf p}$=0 in Eq.\ (\ref{d})],
\begin{equation}
\label{useful}
U^{\dag}\equiv\sum_ka_{k}^{\dag}b_{-k}^{\dag}=N^{1/2}d_{0},
\end{equation}
so that 
$HU^{\dag}|0\rangle=N^{1/2}H_Xd_0^{\dag}|0\rangle=\Omega_0U^{\dag}|0\rangle$,
where $\Omega_0$ is the detuning from the magnetoexciton level.
On the other hand, as we will see later, the {\em nonlinear} optical
response  is dominated  by the exciton-exciton interactions.

Let us now turn to the two-exciton states which govern the FWM signal. 
Only states with zero total
momentum contribute to the optical response. 
The normalized two-exciton basis with zero total momentum
can be chosen as
\begin{equation}
\label{Psi-twoexc}
\Psi_{\bf p}^{(2)}({\bf r}_1,{\bf r}_2;{\bf r}'_1,{\bf r}'_2)
=\Psi_{\bf p}({\bf r}_1,{\bf r}_2)\Psi_{\bf -p}({\bf r}'_1,{\bf r}'_2).
\end{equation}
Correspondingly, the states 
$|{\bf p, -p}\rangle= d_{\bf p}^{\dag}|0\rangle
\times d_{\bf- p}^{\dag}|0\rangle$ 
form a complete basis set of the  zero-momentum 
two-exciton Hilbert subspace. Using Eq.\ (\ref{d}), one can
easily expand any two-exciton state in this basis. For example,
\begin{eqnarray}
\label{expan}
d_0^{\dag}d_0^{\dag}|0\rangle= 
d_0^{\dag}|0\rangle\times d_0^{\dag}|0\rangle
-\frac{1}{N}\sum_{\bf p} d_{\bf p}^{\dag}|0\rangle
\times d_{\bf- p}^{\dag}|0\rangle.
\end{eqnarray}
The matrix elements of the Coulomb potential between
four photoexcited particles, two electrons and two holes, 
evaluated in the two-exciton basis (\ref{Psi-twoexc}), have
the form
\begin{eqnarray}
V_{ee}({\bf p},{\bf q})&=&
v_{|{\bf p}-{\bf q}|}
e^{-({\bf p}-{\bf q})^2l^2/2-i({\bf p}\times{\bf q})\cdot{\bf z}l^2},
\\
V_{hh}({\bf p},{\bf q})&=&
v_{|{\bf p}-{\bf q}|}
e^{-({\bf p}-{\bf q})^2l^2/2+i({\bf p}\times{\bf q})\cdot {\bf z}l^2},
\\
V_{eh}({\bf p},{\bf q})&=&
-2v_{|{\bf p}-{\bf q}|}
e^{-({\bf p}-{\bf q})^2l^2/2}
+2\varepsilon_p\delta_{\bf p q},
\end{eqnarray}
with $\varepsilon_p$ given by Eq.\ (\ref{spectr}). The 
two-exciton matrix elements of the Hamiltonian $H$
are then given by 
\begin{eqnarray}
\label{H-twoexc}
H({\bf p},{\bf q})=2\Omega_p \delta_{\bf pq}
-V({\bf p},{\bf q})/L^2.
\end{eqnarray}
The first term is simply the energy of two non-interacting excitons
(in the rotating frame). 
The second term represents the exciton-exciton interaction potential
\begin{equation}
\label{exc-exc}
V({\bf p},{\bf q})=4v_{|{\bf p}-{\bf q}|}
\sin^2[({\bf p}\times{\bf q})\cdot {\bf z}\, l^2/2] \
e^{-({\bf p}-{\bf q})^2l^2/2} + V_A({\bf p},{\bf q})
=V_S({\bf p},{\bf q})+V_A({\bf p},{\bf q}).
\end{equation}
The potential $V_A({\bf p},{\bf q})$ 
describes the contribution to the exciton-exciton 
interactions in the quasi-2D case that
comes from the lifting of the electron-hole
symmetry in a strong magnetic field by, e.g., the differing band 
offsets, the lateral confinement, 
and the disorder.\cite{bychkov83,chernyak98}
Such an asymmetry can be strong in typical quantum wells and plays
an important role in the nonlinear
optical response in QW's.\cite{chernyak98,qhe}

Note here that the symmetric part of the exciton-exciton potential, 
$V_S({\bf p},{\bf q})$, and the exciton
dispersion are not independent, but are related as follows:
\begin{equation}
\label{ident}
2\varepsilon_p - 2\varepsilon_0= 
\int\frac{d{\bf q}}{(2\pi)^2}V_{S}({\bf p},{\bf q}).
\end{equation}
This relation 
reflects the two possible ways of arranging two electrons and two
holes into two excitons. 

Note also that $V_{S}({\bf p},{\bf q})$ vanishes
for zero momenta. This reflects the fact that,
due to the electron and hole symmetry in the ideal 2D case, 
the net interaction between  two magnetoexcitons  vanishes  at small
{\em e-h} separation (which corresponds to small momenta).  
In contrast, the potential $V_A({\bf p},{\bf q})$, originating from
the breaking of the {\em e-h} symmetry in a QW,\cite{qhe}
is finite for small momenta and 
leads to the main interaction-induced contribution to the 
FWM signal.\cite{chernyak98,yoko99} 

\section{Third-order optical response of 2D magnetoexcitons}
\label{sec:fwm}

As discussed in the Introduction, the calculation of the
magnetoexciton nonlinear optical response in QW's is simplified  
considerably   due to the 
``freezing'' of the kinetic energy of electrons and
holes by strong  perpendicular magnetic field.
Using the canonical transformation method,\cite{perakis99,primozich00}
outlined in Appendix \ref{app:gen}, we have obtained a simple expression for the 
FWM polarization $\tilde{P}_{XX}$ that comes from the exciton-exciton
interactions. Refering the reader to Appendix \ref{app:fwm} for the derivation,
the result reads
\begin{eqnarray}
\label{fwm-final1} 
\tilde{P}_{XX}(t,\tau)=i\mu^2{\cal E}_1 \theta(t+\tau) 
e^{- \Gamma ( t + \tau)-i \omega_0(t-\tau)}
\biggl[e^{-i\Omega_0(t+\tau)}\chi(-\tau) 
- e^{i\Omega_0(t+\tau)}\chi(t) \biggr],
\end{eqnarray} 
Here, ${\cal E}_i(t)$ is the electric field of the probe ($i=1$) or pump
($i=2$), $\mu$ is the interband dipole matrix element, 
$\omega_0$ is the central optical frequency, 
$\Omega_0$  is the pump detuning from the magnetoexciton level,
$\Gamma$ is the magnetoexciton dephasing rate, and $\theta(t)$ is the
step-function.
Since the FWM polarization along the direction
$2 {\bf k}_2-{\bf k}_1$ is linear in the probe 
optical field,  we assume, for simplicity,
a delta-function probe pulse 
${\cal E}_1(t)=e^{-i\omega_0\tau}{\cal E}_1\delta(t+\tau)$
and a pump pulse of finite duration $t_0$ centered at $t=0$. 

As can be seen from Eq.\ (\ref{fwm-final1}), the interaction-induced
FWM polarization is directly proportional 
to the two-exciton amplitude  $\chi(t)$.
In Appendix \ref{app:fwm}  we show that, in the case of
magnetoexcitons in the LLL,  
$\chi(t)$ satisfies a simple equation: 
\begin{eqnarray} 
\label{Eq-chi}
i\partial_t \chi(t)-2\Omega_0\chi(t)=
V^{HF}p^2(t) + F(t),
\end{eqnarray}
where $p(t)$ is the linear pump polarization,
determined by Eq.\ (\ref{eqp1}), and
$V^{HF}$ is the Hartree-Fock interaction
given by Eq.\ (\ref{hartree1}).
The last term in Eq.\ (\ref{Eq-chi}) describes the effect
0f four-particle correlations and has the form 
(see Appendix \ref{app:fwm}) 
\begin{eqnarray} 
\label{corr1}
F(t)=\frac{l^2}{\pi} \int d{\bf q} \, V_q^{HF} \, w_q(t),
\end{eqnarray}
where $w_p(t)$ are the coefficients of the expansion of the two-exciton 
wave-function in the basis (\ref{Psi-twoexc}).
As shown in Appendix \ref{app:fwm} , they satisfy the Wannier-like equation 
\begin{eqnarray} 
\label{eqwXX3}
i\partial_t w_p(t)-2\Omega_pw_p(t)=
-\int\frac{d{\bf q}}{(2\pi)^2}V({\bf p},{\bf q})w_q(t)
+V_p^{HF}p^2(t),
\end{eqnarray}
where
$V_p^{HF}$ 
is the Hartree-Fock potential whose explicit expression 
is given by Eq.\ (\ref{hartree}). The amplitude
$\chi(t)$ is related to $w_p(t)$ simply as
$\chi(t)=w_0(t)-N^{-1}\sum_{\bf p}w_p(t)$ (see Appendix \ref{app:fwm}).
The source (last) term in the rhs of Eq.\ (\ref{eqwXX3})
describes the interaction 
of two pump-induced polarizations, $p^2(t)$, via the 
Hartree-Fock potential $V_p^{HF}$, while the  
first term describes 
the effect of the exciton-exciton correlations.

The Hartree-Fock approximation
corresponds to setting $F(t)=0$ in Eq.\ (\ref{Eq-chi}), 
which yields
\begin{eqnarray} 
\label{chi-hartree}
\chi^{HF}(t)=-iV^{HF}\int_{-\infty}^{t}dt'e^{-2i\Omega_0(t-t')}p^2(t'),
\end{eqnarray}
with the pump polarization $p(t)$ given by the solution of Eq.\ (\ref{eqp1}),
\begin{eqnarray} 
\label{p}
p(t)=-i\mu\int_{-\infty}^{t}dt'e^{-i(\Omega_0-i\Gamma)(t-t')}{\cal E}_2(t').
\end{eqnarray}
The FWM polarization within the Hartree-Fock approximation
is then obtained by substituting $\chi^{HF}(t)$
into Eq.\ (\ref{fwm-final1}).
The relation between the above formalism and that of 
Refs.\ \onlinecite{chemla99,kner99,schafer96} is discussed in Appendix D.

\section{Analytic solution in strong magnetic field}
\label{sec:analytic}

The role of the exciton-exciton correlations in the FWM
polarization is described by the function $F(t)$, given by 
Eq.\ (\ref{corr1}).
In order to obtain $F(t)$, it is necessary to find the two-exciton
amplitude $w_p(t)$ by solving Eq.\ (\ref{eqwXX3}). 
In the general case, this can only be done numerically, and 
the corresponding results are  presented in the next section. 
In this section, we derive an analytical expression for $w_p(t)$ in
the case of strong magnetic field.

As discussed above, the interactions between QW magnetoexcitons,
confined to the LLL,
contribute to the nonlinear optical polarization only in
the presence of electron-hole asymmetry.\cite{chernyak98,yoko99}  The
corresponding potential 
$V_{A}({\bf p},{\bf q})$ is short-ranged, and its specific form
depends on the sample.\cite{chernyak98}  In order to proceed further,
we assume the following form for the s-wave component of 
this potential:
\begin{equation}
\label{VA} 
V_{A}({\bf p},{\bf q}) = V_{A} e^{-(p^2 + q^2) \,  a^{2}/2},
\end{equation} 
where $a$ is the potential range. The potential strength
is characterized by the energy scale
\begin{equation}
\label{u0} 
u_0= \frac{V_A}{4 \pi a^2}.
\end{equation}
With such $V_{A}({\bf p},{\bf q})$,
the   Hartree-Fock parameters, determined by 
Eqs. (\ref{hartree}) and (\ref{hartree1}), have a simple form,
\begin{equation}
\label{hartree3} 
V_p^{HF} = u_0
\Biggl(1-\frac{a^2}{l^2}\Biggr) e^{-p^{2}a^{2}/2}.
\end{equation}
and 
\begin{equation}
\label{hartree4}  
V^{HF} = -u_0
\Biggl(\frac{l}{a}-\frac{a}{l}\Biggr)^2.
\end{equation}
Turning to Eq.\ (\ref{eqwXX3}), 
We observe that a considerable simplification occurs
for sufficiently strong magnetic field such that  $l<a$.
In this case, the characteristic momenta of the excitons
scattered by the 
potential (\ref{VA}) are small: 
$p\sim a^{-1} < l^{-1}$. Since for small momenta the
magnetoexciton size is also small (see Section II), 
the {\em symmetric} part of the exciton-exciton potential, 
$V_S({\bf p},{\bf q})$, is suppressed. Indeed, for
the characteristic momenta $p,q\sim 1/a$, we get from 
Eq.\ (\ref{exc-exc}) that 
$V_S({\bf p},{\bf q})\sim  e^2 a (l/a)^4$. 
Therefore, $V_S({\bf p},{\bf q})$ can be
neglected, as compared to $V_A({\bf p},{\bf q})$, under the
condition 
\begin{equation}
\label{condition}
u_0>E_0 \Biggl(\frac{l}{a}\Biggr)^5,
\end{equation}
where $E_0\sim e^2/\kappa l$ is the magnetoexciton binding energy 
[see Eq.\ (\ref{spectrum})]. Note that the above condition can be  met
even for weak asymmetry, $u_0 \ll  E_0$, provided that the 
magnetic field is sufficiently strong. 
Thus, for $l<a$, we can
replace $V({\bf p},{\bf q})$ by $V_A({\bf p},{\bf q})$
in the first term in the rhs of Eq.\ (\ref{eqwXX3}).
Then, by using   Eqs. (\ref{corr1}), (\ref{VA}), and (\ref{hartree3}),
this term can be expressed via  $F(t)$ as
\begin{eqnarray} 
-\int\frac{d{\bf q}}{(2\pi)^2}V_A({\bf p},{\bf q})w_q(t)
=-\frac{l^2 \ e^{-p^{2} a^{2}/2}}{\pi(l^2/a^2-1)} \, 
\int \, d{\bf q} \, V_q^{HF} \, w_q(t)
=-F(t) \, \frac{e^{-p^{2} a^{2}/2}}{l^2/a^2-1}.
\end{eqnarray} 
Eq.\ (\ref{eqwXX3}) can be solved by Fourier transform:
\begin{eqnarray}
\label{WXX-fourier}  
w_p(\omega)=-\frac{e^{-p^2a^2/2}}{l^2/a^2-1} \, 
\frac{F(\omega)}{\omega-2\Omega_p}
+\frac{V_p^{HF}p^{(2)}(\omega)}{\omega-2\Omega_p},
\end{eqnarray} 
where $p^{(2)}(\omega)$ is the Fourier transform of $p^2(t)$.
To find $F(\omega)$, we 
multiply the above equation by $V_p^{HF}$ and take the sum over 
${\bf p}$. This gives  
\begin{eqnarray}
\label{eqF}
F(\omega)=-\frac{l^2}{\pi} \frac{F(\omega)}{l^2/a^2-1}
\int d{\bf p} \frac{e^{-p^2a^2/2} V_p^{HF}}{\omega-2\Omega_p}
+p^{(2)} (\omega)  \frac{l^2}{\pi} \int d{\bf p}
\frac{(V_p^{HF})^2}{\omega-2\Omega_p}, 
\end{eqnarray} 
yielding
\begin{eqnarray}
\label{F-fourier}  
F(\omega)=-V^{HF} p^{(2)}(\omega) \frac{Q(\omega)}{1+Q(\omega)},
\end{eqnarray} 
where
\begin{equation}
\label{Q}
Q(\omega)=\frac{V_A}{(2\pi)^2} \int d{\bf p}
\frac{ e^{- p^{2} a^{2}}}{\omega-2\Omega_p}.
\end{equation}
Using Eq.\ (\ref{F-fourier}), the amplitude $\chi(\omega)$ can then be
easily obtained from Eq.\ (\ref{Eq-chi}) as 
\begin{equation}
\label{chi-anal}
\chi(\omega)
=\frac{V^{HF} \ p^{(2)}(\omega)}{\omega-2\Omega_0} \ S(\omega),
\end{equation}
where 
\begin{equation}
\label{S}
S(\omega)=\frac{1}{1+Q(\omega)}.
\end{equation}
Equations (\ref{Q}-\ref{S}), which are the main result 
of this section, 
provide an analytic expression for the two-exciton amplitude
$\chi(t)$, which determines the FWM
polarization (\ref{fwm-final1}). The result (\ref{chi-anal}) has the
form of a product of the Hartree-Fock result,
Eq.\ (\ref{chi-hartree}),  and  the {\em frequency-dependent} 
factor $S(\omega)$:
$\chi(\omega)=\chi^{HF}(\omega)S(\omega)$.  In the time domain,
$\chi(t)$ has the simple form
\begin{equation}
\label{chi-final}
\chi(t)=\int_{-\infty}^{\infty}dt'S(t-t')\chi^{HF}(t')
\end{equation}
with $\chi^{HF}(t)$ given by  Eq.\ (\ref{chi-hartree}). As can be seen, 
$S(t)$ represents a  {\em memory kernel} whose time dependence is
governed  by the exciton-exciton correlations.
In the absence of correlations, corresponding to $Q(\omega)=0$ in
Eq.\ (\ref{S}), the kernel is instantaneous, $S(t)=\delta(t)$, and
we recover the Hartree-Fock result.

\section{Memory function}
\label{sec:memory}

In this Section we analyze the properties 
of the memory function 
$S(\omega)$, given by Eq.\ (\ref{S}).
For  sufficiently strong magnetic fields 
such that $l < a$,  the main  contribution 
to the integral in Eq.\ (\ref{Q}) comes from   
small momenta $q\sim a^{-1}< l^{-1}$. 
For such momenta, the exciton dispersion is quadratic, 
$\Omega_q\simeq \Omega_0+q^2/2M$, 
where $M$ is magnetoexciton effective 
mass [see Eq.\ (\ref{spectrum})]. We then get from Eq.\ (\ref{Q})
\begin{equation} 
\label{Q1}
Q(\omega) =V_A{\cal N} \int_0^{\infty} dE
\frac{e^{-E/D}}{\omega-2\Omega_0-2E},
\end{equation}
where ${\cal N}=M/2\pi=\frac{4}{3\pi}(E_0l^2)^{-1}$
is the 2D {\em magnetoexciton} density of states and 
\begin{equation}
\label{D} 
D=\frac{1}{2 M a^2}=\frac{3E_0}{16}\Biggl(\frac{l}{a}\Biggr)^2
\end{equation} 
plays the role of the cutoff energy.

For large frequencies, 
$|\omega-2\Omega_0| \gg D$, the integrand in  
Eq.\ (\ref{Q1}) can be expanded in  terms of 
$(\omega - 2 \Omega_0)^{-1}$ yielding
\begin{equation} 
\label{hf} 
Q(\omega) \simeq \frac{ u_0}{ \omega - 2 \Omega_0}.
\end{equation} 
For small frequencies, $|\omega-2\Omega_0| \ll D$,
the integral in Eq.\ (\ref{Q1}) diverges at the lower
limit and  can be estimated as
\begin{equation}
\label{Q-anal}
Q(\omega)\simeq - G \ln \left| \frac{2D}{\omega-2\Omega_0} \right| 
- i \pi G \theta(\omega-2\Omega_0),
\end{equation} 
where
\begin{equation} 
\label{G} 
G=\frac{V_A{\cal N}}{2}=\frac{8}{3}\frac{u_0}{E_0}
\Biggl(\frac{a}{l}\Biggl)^2=\frac{u_0}{2D}
\end{equation}
is a dimensionless parameter characterizing the
strength of the exciton-exciton interaction. 
As can be seen from Eq.\ (\ref{G}), 
the magnitude of $G$
can be tuned by varying the magnetic field:
$G$ increases with  decreasing  the 
magnetic length $l$.
Note also that the logarithmic frequency dependence of 
$Q(\omega)$ is similar to that of 
the exciton-exciton scattering matrix.\cite{bychkov83,chernyak98}

In the following,  we will distinguish 
between attractive and repulsive exciton-exciton interactions.

\subsection{Attractive interaction}

We first 
consider the case of attractive exciton-exciton interactions,
corresponding to $G>0$. 
For $\omega<2\Omega_0$, the imaginary part of 
$Q(\omega)$ vanishes and the memory function $S(\omega)$ 
exhibits a biexciton pole. In the case of weak 
exciton-exciton interaction (weak asymmetry), the corresponding
frequency $\omega_b$ can be determined by using the low-frequency
asymptotics (\ref{Q-anal}) of $Q(\omega)$:
$1- G\ln \left|\frac{2D}{\omega_b-2\Omega_0}\right|=0$.
This gives 
\begin{equation}
\label{biexc}
\omega_b=2\Omega_0-E_b,
~~~
E_b=2De^{-1/G},
\end{equation}
where $E_b$ is the biexciton binding energy. 
Note that the above expression holds
for $G \ll 1$ corresponding to small biexciton energy $ E_{b} \ll D$.
The biexciton contribution to
$S(\omega)$ can be obtained by expanding 
Eq.\ (\ref{Q-anal})
in the vicinity of $\omega_b$:
\begin{equation}
\label{S-biexc}
S_b(\omega)\sim-\frac{E_b}{G}\frac{1}{\omega-2\Omega_0+E_b+i\gamma_b},
\end{equation}
or, in the time domain,
\begin{equation}
\label{S(t)}
S_b(t)= \frac{ i E_b}{G} \theta(t)e^{i (E_{b} - 2 \Omega_0)t- \gamma_b t}, 
\end{equation} 
where $\theta(t)$ is the step function
and we have included the biexciton width $\gamma_b$ 
that mainly comes from electron-phonon processes not incorporated in the
Hamiltonian $H$ considered here.

In the time domain, the biexciton resonance leads to 
oscillations of the memory kernel 
with a period determined by 
the biexciton binding energy
(see Eq.\ (\ref{S(t)})).  The memory effects 
decay on a time scale determined by the {\em biexciton}  
dephasing time, which, in a magnetic field, can be  considerably
{\em longer} 
than the exciton dephasing time.\cite{kner98,kner99}
Near the biexciton resonance, the amplitude $\chi(\omega)$ 
[see Eq.\ (\ref{chi-anal})] that determines the FWM 
signal, takes the form 
\begin{equation}
\label{biexc1}
\chi_b(\omega)=G^{-1} \
\frac{V^{HF}p^{(2)}(\omega)}{\omega-2\Omega_0+E_b+i\gamma_b},
\end{equation}
and exhibits a biexciton pole as a function of frequency.

The  memory function $S(\omega)$
also exhibits a second resonance centered at the frequency 
$\omega_{sc}=2\Omega_0+E_b$,
which describes the  continuum band of exciton-exciton scattering states.
The width of this {broad continuum resonance is determined by
the exciton-exciton interactions. 
Indeed, for $\omega>2\Omega_0$,  the 
function $Q(\omega)$, that describes the exciton-exciton
scattering, develops an imaginary part 
proportional to the interaction strength [see Eq.\ (\ref{Q-anal})].
For $E_b > \gamma_b$, the scattering band and biexciton
resonance are well separated from each other
and lead to distinct dynamics, 
while for $E_b < \gamma_b$ they merge 
into a single asymmetric peak.

The above results hold in the case of weak exciton-exciton
scattering, $G \ll 1$. With increasing magnetic 
field,  the parameter $G$, Eq.\ (\ref{G}), 
that characterizes the strength of the exciton-exciton 
interaction increases, and so does the 
biexciton binding energy $E_b$. 
For sufficiently strong magnetic fields 
such that $E_b \gg D$, corresponding to $G\gg 1$, 
one can use the high  
frequency asymptotic expansion of 
$Q(\omega)$, Eq.\ (\ref{hf}), which 
when substituted 
into Eqs.\ (\ref{S}) and (\ref{chi-anal})
gives 
\begin{equation}
\label{GAPM}
\chi(\omega)=
\frac{V^{HF}p^{(2)}(\omega)}{\omega-2\Omega_0+u_0+i\gamma_b}.
\end{equation}
Note that for $G\sim 1$, we have $E_b \sim u_0$, and 
Eqs.\ (\ref{GAPM}) and \ (\ref{biexc1}) match.

It is important to note that Eq.\ (\ref{GAPM}) 
becomes exact in the strong field limit, $l \ll a$. 
Indeed, in this case,
the typical magnetoexciton  momenta are small, 
$q\sim a^{-1}\ll l^{-1}$,  so that one can disregard
the magnetoexciton dispersion in 
Eq.\ (\ref{Q}) and replace $\Omega_q$ by $\Omega_0$, 
which leads again to  Eq.\ (\ref{hf}) 
for $Q(\omega)$ and, therefore, to
Eq. \ (\ref{GAPM}) for $\chi(\omega)$.
In other words, for $G\gg 1$, the exciton-exciton scattering is 
suppressed and the memory effects are dominated by the biexciton
resonance. 
The simple expression (\ref{GAPM}) can be then viewed as a 
``biexciton pole approximation'' for the two-exciton amplitude 
$\chi(\omega)$. In fact, Eq.\ (\ref{GAPM}) is equivalent to the  
average polarization model\cite{wegener90,ssr91,schafer96,chemla99,kner99}
(APM), as we show in Appendix \ref{app:APM}.

In Fig.\ (\ref{fig:1}) we plot the memory function,
obtained  by evaluating Eqs.\ (\ref{S}) and (\ref{Q}) with the full
exciton dispersion (\ref{spectr}).
As we can see, for attractive iteraction, $S(\omega)$ is mainly
dominated by the biexciton pole. 

\subsection{Repulsive interaction}

Let us now turn to the case of repulsive exciton-exciton interactions,
$G<0$. 
In this case, the memory function describes  a continuum
band of scattering states above the two-exciton energy.
We start with the case of 
strong interactions $|G|\gg 1$.
Here we can use again the high frequency asymptotics (\ref{hf}) for
${\rm Re}Q(\omega)$
(valid for $\omega-2\Omega_0 \gg D$).
We see from Eq.\ (\ref{S}) that
$S(\omega)$ exhibits a resonance 
at $\omega =2\Omega_0 + |u_0|$. 
In contrast to the attractive case however, in this frequency
range $Q(\omega)$ develops imaginary part, which can be deduced 
form Eq.\ (\ref{Q1}) as
${\rm Im}\,Q(\omega)=
-i\pi Ge^{-(\omega-2\Omega_0)/2D}\theta(\omega-2\Omega_0)$. 
Near the resonance,  we obtain from Eqs.\ (\ref{S}) and 
Eq.\ (\ref{chi-anal}) that
\begin{equation}
\label{S-repul}
\chi(\omega)=\frac{V^{HF}p^{(2)}(\omega)}
{\omega-2\Omega_0-|u_0|+i\gamma_{sc}}, 
~~~\gamma_{sc}=\pi u_0Ge^{-|G|}.
\end{equation}
The above expression describes a resonance made up from the
continuum of scattering states. Eq.\ (\ref{S-repul}) is valid for
$|u_0|/2D=|G|\gg 1$, indicating that in this case the resonance is
sharp. Comparing to Eq.\ (\ref{S-repul}), we can view this continuum
resonance as an antibound state. The crucial difference, however, is that 
here the resonance width $\gamma_{sc}$ is determined by the
exciton-exciton interactions rather than homogeneous broadening due to
phonons. The narrow width of the resonance comes from the fact that
for $|G|\gg 1$, the momentum exchange processes between excitons are
suppressed. Note here that the APM (\ref{GAPM}) is equivalent to 
Eq.\ (\ref{S-repul}) but with   $\gamma_{sc}$ introduced phenomenologically. 

With decreasing $|G|$, the lineshape of the memory function
changes. This is illustrated in  Fig.\ (\ref{fig:2}). 
The sharp Lorentzian peak transforms into a broad asymmetric band
centered at lower frequencies. Importantly, $S(\omega)$ develops a
{\em cusp} at $\omega=2 \Omega_0$. This cusp is manifestation of the
logarithmic singularity in the low-frequency asymptotics of $Q(\omega)$
[see Eq.\ (\ref{Q-anal})].
In the next section we will see that this 
low-frequency behavior of the memory function has a dramatic effect
on the FWM polarization for long time delays.

\section{Numerical results and  discussion}
\label{sec:numeric}

In this section, we present the results of 
numerical calculations of the third-order FWM polarization
and compare them to the analytical solution of 
Sec.\ \ref{sec:analytic} and \ref{sec:memory}.
The calculations below also incorporate the Pauli blocking
contribution, 
$\tilde{P}_{PB}(t,\tau)$, to the total FWM polarization  
$\tilde{P}(t,\tau) = \tilde{P}_{XX}(t,\tau)
+\tilde{P}_{PB}(t,\tau)$, which is given in Appendix \ref{app:PB}.
We use the standard
conventions according to which the positive time-delay, $\tau>0$,
corresponds to probe pulse arriving before the pump pulse.
We will be interested mainly in the 
time-integrated and spectrally-resolved FWM signals, 
defined as
\begin{eqnarray}
\label{TI-FWM}
\mbox{TI-FWM}=\int_{-\infty}^{\infty} dt|\tilde{P}(t,\tau)|^2
\end{eqnarray}
and
\begin{eqnarray}
\label{SR-FWM}
\mbox{SR-FWM}=|\tilde{P}(\omega,\tau)|^2,
\end{eqnarray}
respectively, where $\tilde{P}(\omega,\tau)$ is the 
Fourier transform of $\tilde{P}(t,\tau)$.

\subsection{Attractive interaction}

We start with the case of attractive exciton-exciton interactions,
$u_0>0$. In Fig.\ \ref{fig:3} we show the results of our numerical
calculation of the TI-FWM signal for a strong magnetic field,
$a/l=6.0$,  and  for both strong and weak {\em e-h} asymmetry.
In order to be consistent with 
the recent experimental data,\cite{kner98,kner99} we have chosen the
biexciton homogeneous broadening, $\gamma_b$, to be 
smaller than that of the exciton, $\Gamma$. 
In both cases, the calculations
yield a strong TI-FWM signal at negative time delays.
For strong asymmetry (large $u_0$),  
the biexciton oscillator strength in the memory function  
$S(\omega)$ is enhanced (see Fig.\ (\ref{fig:1})), 
and the TI-FWM signal exhibits biexciton
oscillations. This behavior can be easily deduced 
from the analytic results of the previous section. 
Indeed, for $\delta$-function pump pulse centered at $t=0$, 
the pump polarization (\ref{p}) is given by
$p(t)=-i\mu {\cal E}_2\theta(t)e^{-i\Omega_0 t-\Gamma t}$, and 
the biexciton contribution to the amplitude $\chi(t)$ 
[Eq.\ (\ref{biexc1})] takes the  simple form
\begin{eqnarray}
\label{biexc2}
\chi_b(t)=\frac{\theta(t)e^{-2i\Omega_0 t}(\mu {\cal E}_2)^2}
{G[E_b-i(2\Gamma -\gamma_b)]}
\Biggl(e^{iE_bt-\gamma_b t}-e^{-2\Gamma t}\Biggr).
\end{eqnarray}
The first term in the rhs leads to the oscillations in the TI-FWM
signal with the period $T_b=2\pi/E_b$, and to the 
overall exponential decay of the  TI-FWM signal with the
characteristic time $\gamma_b^{-1}$ at long negative time
delays. Such a behavior is also  consistent with the calculations 
in Ref. \onlinecite{yoko99}.

In Fig.\ \ref{fig:4}, we plot the 
time evolution of the corresponding  SR-FWM signal, 
which shows both an 
exciton and a (weaker) biexciton peak. 
A distinguishing feature of the spectra is that
the height of the {\em exciton} peak oscillates as a
function of time delay with a period determined by the 
{\em biexciton} binding energy, $T_b=2\pi/E_b$. This behavior is
easily reproduced by using  Eq.\ (\ref{biexc2}) (see below).

As mentioned in the previous section, in the 
strong magnetic field
case $l/a\ll 1$, the exciton-exciton scattering is
suppressed and the FWM polarization is governed by the biexcitonic
effects. In  Fig.\ \ref{fig:5}, we show the 
TI-FWM signal for weaker magnetic 
field such that  $a/l=3.0$ and for two different 
pump durations, both of which 
are shorter than the exciton dephasing time, 
$t_0 < \Gamma^{-1}$.
In Fig.\ \ref{fig:5}(a),
the pulse duration was chosen to be shorter than the 
inverse biexciton energy, $t_{0} < E_{b}^{-1}$. In this case,
the TI-FWM curves are qualitatively  similar to those in 
Fig.\ \ref{fig:3}(a); in particular, they 
exhibit an overall exponential decay at long negative time delays
together with biexciton oscillations as $u_0$ increases. 
In contrast, for the longer pulse duration
$t_{0}> E_{b}^{-1}$, the change in the shape of the TI-FWM
signal with increasing $u_0$ is different 
[see Fig.\ \ref{fig:5}(b)].

To understand this behavior, we note that, for weak asymmetry
$u_0/E_0\ll 1$,  we have $t_{0} < E_{b}^{-1}$, and the  decay of the
TI-FWM signal is similar to that for the shorter pulse duration 
[compare Figs.\ \ref{fig:5}(b) and \ \ref{fig:5}(a)].
On the other hand, for larger $u_0$ we have $t_{0} > E_{b}^{-1}$, 
and the pump pulse tuned at the magnetoexciton level, $\Omega_0=0$,
does not directly excite the biexciton.
In this case the decay at intermediate times becomes 
non-exponential , while the amplitude of the oscillations is
significantly enhanced [see Fig.\ \ref{fig:5}(b)].
The reason for such a behavior is that, 
for short time delays, the signal 
decays exponentially with the Hartree-Fock decay
time\cite{wegener90} 
$(4\Gamma)^{-1}$ [see Fig.\ \ref{fig:3}(b)], while for longer time delays,
the build-up of the biexciton correlations leads to an exponential
decay with a  different characteristic time,  $(2\gamma_b)^{-1}$, 
determined by the biexciton width. The crossover between these two
regimes leads to an apparent nonexponential decay at intermediate
time delays [see Fig.\ \ref{fig:3}(b)].

The sensitivity to the detuning of the time-evolution of the FWM
polarization becomes even more evident in the SR-FWM
signal. In Figs.\ \ref{fig:6} and \ref{fig:7}, we plot the
corresponding spectra versus time delay for pump tuned to the 
exciton and biexciton levels, respectively. 
For $\Omega_0=0$, the exciton peak decays
exponentially for positive time delays, $\tau>0$, with 
characteristic time $(2\Gamma)^{-1}$, while the biexciton peak is
suppressed (see Fig.\ \ref{fig:6}).
For pump tuned at the biexciton, $2\Omega_0=E_b$, the situation changes:
for $\tau > 0$ the decay of the exciton  peak
follows the pump pulse, while the
biexciton peak decays exponentially with a 
characteristic time $ (2\Gamma)^{-1}$ (see Fig.\ \ref{fig:7}). 
For negative time delays, the exciton peak oscillates with the period
$2\pi/E_b$ (compare to Fig.\ \ref{fig:4}).

To understand this behavior, consider the
biexciton contribution to the frequency-dependent polarization  
$\tilde{P}_{XX}(\omega,\tau)$.
The latter can be presented as a sum of two terms 
$\tilde{P}_b^{(1)}(\omega,\tau)+\tilde{P}_b^{(2)}(\omega,\tau)$,
corresponding to the first and second terms in  Eq.\ (\ref{fwm-final1}).
The first term can be presented as 
\begin{eqnarray}
\label{SR1}
\tilde{P}_b^{(1)}(\omega,\tau)=
-\frac{\mu^2 {\cal E}_1e^{i(\bar{\omega}+\Omega_0)\tau} \chi_b(-\tau)}
{\bar{\omega}+i\Gamma},
\end{eqnarray}
where $\bar{\omega}=\omega-\bar{E}_g+E_0$ is the frequency measured
from the magnetoexciton level 
(we used here that $\Omega_0=\bar{E}_g-E_0-\omega_0$). 
Equation (\ref{SR1}) describes the
peak at the exciton energy, $\bar{\omega}=0$. The peak height depends
on the time delay via $\chi_b(-\tau)$. The latter, for short pulse
duration, can be approximated by  Eq.\ (\ref{biexc2}), 
leading, for $\tau<0$,  to the oscillations of the exciton peak
amplitude with the period $2\pi/E_b$ determined by the biexciton
binding energy (see Figs.\ \ref{fig:4} and \ref{fig:5}).
Note that, since $\chi_b(t)=0$ for times $t<0$ prior to the onset of
the pump pulse [see Eq.\ (\ref{biexc2})], the above contribution
vanishes for $\tau > 0$. Therefore, for positive time delays, the 
SR-FWM signal at the exciton frequency is  primarily determined  
by the Pauli blocking term. 

The second term in  Eq.\ (\ref{fwm-final1})
has the form
\begin{eqnarray}
\label{SR2}
\tilde{P}_b^{(2)}(\omega,\tau)=
\mu^2 {\cal E}_1e^{-i(\bar{\omega}+\Omega_0)\tau}
\int_{-\infty}^{\infty}\frac{d\omega'}{2\pi}
\frac{e^{i\omega'\tau} \chi_b(\omega')}
{\bar{\omega}+2\Omega_0-\omega' +i\Gamma}.
\end{eqnarray}
Using  Eq.\ (\ref{biexc1}), the integrand of Eq.\ (\ref{SR2}) can be
decomposed  as 
\begin{equation}
\label{integrand}
\frac{\chi_b(\omega')}
{\bar{\omega}+2\Omega_0-\omega' +i\Gamma}=
\frac{G^{-1}V^{HF}p^{(2)}(\omega')}{\bar{\omega}+E_b+i(\Gamma+\gamma_b)}
\Biggl[
\frac{1}{\omega'-2\Omega_0+E_b+i\gamma_b}
+\frac{1}{\bar{\omega}+2\Omega_0-\omega'+i\Gamma}
\Biggr],
\end{equation}
yielding
\begin{equation}
\label{pol-biexc}
\tilde{P}_b^{(2)}(\omega,\tau)=
\frac{\mu^2 {\cal E}_1G^{-1}e^{-i(\omega+\Omega_0)\tau}}
{\bar{\omega}+E_b+i(\Gamma+\gamma_b)}
\int_{-\infty}^{\infty}\frac{d\omega'}{2\pi}
\Biggl[
\frac{V^{HF}p^{(2)}(\omega')e^{i\omega'\tau}}
{\omega'-2\Omega_0+E_b+i\gamma_b}
+\frac{V^{HF}p^{(2)}(\omega')e^{i\omega'\tau}}
{\bar{\omega}+2\Omega_0-\omega'+i\Gamma}
\Biggr].
\end{equation}
The prefactor in the rhs  of Eq.\ (\ref{pol-biexc})
has a resonance  at the biexciton
frequency, $\bar{\omega}=-E_b$, with width $\Gamma+\gamma_b$.
The amplitude of the biexcitonic peak is 
determined by the integral 
in Eq.\ (\ref{pol-biexc}).
For negative delays, the main contribution
to the above integral  comes from the first term
in the integrand,  whose denominator has a pole 
at $\omega'=2\Omega_0 - E_b$. 
For  resonant pump excitation,  $\Omega_0=0$, 
and pulse duration $t_{0} < \Gamma^{-1}$,
the frequency width of 
$p^{(2)}(\omega')$ in the integrand of Eq.\ (\ref{pol-biexc})
is of the order of  $ 2 \Gamma$. 
In this case, the  contribution of the pole is 
suppressed for $E_b > \Gamma$,
and therefore the  biexciton peak is diminished (see Fig. \ref{fig:6}). 
For off-resonant pump tuned at the biexciton, 
$2\Omega_0\simeq E_b$, and for pulse duration 
$t_{0} > 2/E_{b}\sim \Omega_0^{-1}$, 
the time dependence of $p(t)$ 
is mainly determined by that of the pulse itself,
so that the  frequency 
width of  $p^{(2)}(\omega')$ is  $\sim t_0^{-1}$. 
In this case the pole dominates the integral,  
leading to a strong biexciton peak  in the SR-FWM
signal, which decays on a time scale 
$(2\gamma_b)^{-1}$ (see Fig. \ref{fig:7}). 

For positive delays, $\tau>0$, the main contribution 
to the biexciton term 
$\tilde{P}_b^{(2)}(\omega,\tau)$ 
comes from the second term of the integrand in Eq. (\ref{pol-biexc}),
which exhibits a  pole at 
$\omega'=\bar{\omega} +2\Omega_0$. Since 
$\bar{\omega} \sim -E_b$ due to the prefactor in 
Eq. (\ref{pol-biexc}),  for resonant excitation,
$\Omega_0=0$, the biexciton peak is diminished
for $E_b > \Gamma$
(see Fig. \ref{fig:6}). For off-resonant excitation, 
$2\Omega_0=E_b$, this pole dominates the integral leading to a strong
SR-FWM signal at the biexciton energy
(see Fig. \ref{fig:7}), which decays on a 
time scale 
$(2\Gamma)^{-1}$ [see  Eq.\ (\ref{pol-biexc})]. 
Note that, for pump excitation at the {\em biexciton} 
frequency, $2\Omega_0=E_b$,
the SR-FWM signal  at the {\em exciton} frequency 
is strong  for negative 
time delays due to the exciton-exciton correlations
[see Eq.\ (\ref{SR1})], but diminishes
after the pump pulse is gone due to the suppression 
of both the  Pauli blocking contribution 
and $\tilde{P}_b^{(1)}$ [see Eq.\ (\ref{SR1})].

It should be noted
that for all the above parameter values, the analytical curves
calculated from Eqs.\ (\ref{Q}-\ref{S}), which were derived 
in the previous section for $l < a$,  are 
practically indistiguishable from the  exact numerical calculations in 
Figs.\ \ref{fig:3}-\ref{fig:7}. In fact, Eqs. (\ref{Q}-\ref{S})
provide a very good approximation even for $a\sim l$. 
This is demonstrated in Fig.\ \ref{fig:8}, where 
we present the results of numerical calculations
of the TI-FWM signal at $a/l=1.5$ together with the analytic solution
(\ref{Q}-\ref{S}) for different values of the biexciton width 
$\gamma_b$. For comparison, we also plot the TI-FWM signal derived
using the APM (\ref{GAPM}). 
Note that, for the parameters used in Fig.\ \ref{fig:8}, 
$u_{0}/D \sim 5.0$ and we are therefore  in a regime 
where Eq.\ (\ref{GAPM}) is meaningful.
We see that our analytic
solution  (\ref{chi-anal}), which includes
the effects of the  exciton-exciton scattering
processes, neglected by the APM, 
is in a much better agreement with the 
numerical results even for strong {\em e-h} asymmetry.
In particular, the APM (\ref{GAPM})
fails to provide the correct period and amplitude of the biexciton
oscillations for both $\gamma_b\ll \Gamma$ or 
$\gamma_b=2\Gamma$. In addition, the APM gives a considerably weaker  
magnitude of the FWM signal.

\subsection{Repulsive interaction}

Let us now turn to repulsive exciton-exciton interactions,
$u_0<0$. 
In Fig.\ \ref{fig:9}, we show the TI-FWM signal in a strong
magnetic field, $a/l=3.0$, for several values of 
the parameter $G$
[Eq.\ (\ref{G})] 
characterizing  the strength 
of the exciton-exciton interactions.
For large $|G|$, the signal decays exponentially with a  
characteristic time that increases with $|G|$. In this regime, the
memory function $S(\omega)$ is dominated by an
antibound state well
separated from the exciton resonance (see Fig.\ \ref{fig:2}). 
A narrow  antibound resonance leads to
oscillations in the TI-FWM  signal, similar to biexciton oscillations.
Note that antibound states  were recently observed
experimentally in the FWM signal 
of QW's at zero magnetic field.\cite{nurmikko98}
The decay time is simply the inverse antiresonance width, given
by Eq.\ (\ref{S-repul}). 
Importantly, the latter is determined by
the exciton-exciton interactions, 
consistent with   Fig.\ \ref{fig:9}.

As  $|G|$ decreases, we observe  a transition
from an exponential regime for initial time delays, 
where the antiresonance dominates the decay,  
to a {\em nonexponential} regime for long negative time delays.
For small $|G|$, the latter regime extends down to times $\sim 2/\Gamma$;
for longer times, the TI-FWM signal curves 
up with increasing $|\tau|$ (see Fig.\ \ref{fig:9}).

The long nonexponential tail of the TI-FWM signal has its origin in
the low-frequency behavior of the memory function $S(\omega)$.
As can be seen from  Eq.\ (\ref{fwm-final1}),
the shape of the signal for long negative delays is determined by
the long-time asymptotics of the amplitude $\chi(t)$. The 
latter can be derived by using the analytic expression
(\ref{chi-anal}) for $\chi(\omega)$. For sufficiently large $t$, we
can substitute into Eq.\ (\ref{chi-anal}) 
the low-frequency asymptotics  of
the memory function $S(\omega)$  [see Eq.\ (\ref{Q-anal})],
\begin{equation}
\label{S-anal1}
S(\omega)\simeq \frac{1}
{1+ |G| \ln \left| \frac{2D}{\omega-2\Omega_0} \right| 
+i \pi |G| \theta(\omega-2\Omega_0)}.
\end{equation} 
As shown in Appendix \ref{app:tail}, for short pump
duration this gives 
\begin{equation}
\label{chi-asympt}
\chi(t)\simeq \frac{i(\mu {\cal E}_2)^2}{2\Gamma}
\frac{e^{-2i\Omega_0t}V^{HF}}{1+|G|\ln 2Dt},
\end{equation} 
with $D$ given by Eq.\ (\ref{D}). The above result is valid for 
long times
$t\gg D^{-1}$. The slow logarithmic decay of the amplitude $\chi(t)$
reflects the singular behavior of $S(\omega)$ at small frequencies. 
Note that {\em both} the real and the imaginary part
of the memory function (\ref{S-anal1}) contribute to the
asymptotics (\ref{chi-asympt}) (see Appendix \ref{app:tail}).

Using  Eqs.\ (\ref{chi-asympt}) and (\ref{fwm-final1}), the TI-FWM
signal (\ref{TI-FWM}) for negative time delay takes the form 

\begin{eqnarray}
\label{TI-FWM1}
\mbox{TI-FWM}=\frac{\mu^6{\cal E}_2^4{\cal E}_1^2}{4\Gamma^2}
V^{HF}\int_{|\tau|}^{\infty} dt e^{-2\Gamma (t-|\tau|)}
\Biggl(\frac{1}{1+|G|\ln 2D|\tau|}-\frac{1}{1+|G|\ln 2Dt}\Biggr)^2.
\end{eqnarray}
The main contribution to the integral comes from times 
$(t-|\tau|)\lesssim\Gamma^{-1}$. For $|\tau|\gg \Gamma^{-1}$, the second
term in the rhs of Eq.\ (\ref{TI-FWM1}) can be expanded in
$(t-|\tau|)/|\tau|$, and we finally obtain
\begin{equation}
\label{TI-FWM-asympt}
\mbox{TI-FWM} \propto  \frac{G^2}{\Gamma^5}
\frac{(V^{HF})^2}{\tau^2(1 + |G|\ln 2D|\tau|)^4}.
\end{equation}
The above expression describes the long tail of the TI-FWM signal for
negative time delays in Fig.\ \ref{fig:9}. 
Note that Eq.\ (\ref{TI-FWM-asympt})
is valid for $|\tau|\gg \Gamma^{-1}$, so that with
increasing $\Gamma$, the crossover to the asymptotic behavior 
(\ref{TI-FWM-asympt}) occurs at earlier time delays, as can be seen in
Fig.\ \ref{fig:9}(b).

The behavior of the TI-FWM signal on various time scales 
can be tuned by changing 
the magnetic field strength.
For strong magnetic fields $l \ll a$, the momentum exchange processes
between excitons are suppressed and 
the TI-FWM signal is dominated by the antibound  
state resonance leading to the exponential decay. 
With decreasing magnetic field, 
the energy scale $D$ which determines the characteristic
exciton momenta $q\sim (2MD)^{1/2}$, 
increases [see Eq.\ (\ref{D})], and the 
crossover to the above nonexponential regime
occurs at earlier times $|\tau| > D^{-1}$}
[see Figs.\ \ref{fig:10}(a) and \ref{fig:11}(a)].
At the same time, the shape of the TI-FWM signal
for $|\tau| > D^{-1}$
depends strongly on G. 
For $|G|\sim 1$, the decay is described by the equation 
\begin{equation} 
\label{TI-FWM-asympt1}
\mbox{TI-FWM} \propto  \frac{(V^{HF})^2}{\Gamma^5G^2}
\frac{1}{\tau^2(\ln 2D|\tau|)^4}.
\end{equation}
For $|G|\ll 1$, the situation is more intricate. In this case, the 
asymptotics (\ref{TI-FWM-asympt1}) applies for very long time delays,
$|\tau|\gtrsim D^{-1}e^{\frac{1}{|G|}}\gg D^{-1}$, while for 
$D^{-1}\lesssim|\tau|\lesssim D^{-1}e^{\frac{1}{|G|}}$
the decay is quadratic. 

The behavior of the TI-FWM signal at initial times,
$|\tau|\lesssim D^{-1}$, is
mainly determined by the imaginary part of $S(\omega)$. 
At small $|G|$, the
memory function is a smooth function of  frequency for 
$\omega\gtrsim D$
(see Fig.\ \ref{fig:2}), and, correspondingly,
$S(t)$ is nearly instantaneous, so that the TI-FWM signal shows a
Hartree-Fock-like decay 
(see curves with $|G|=0.24$ in  Fig.\ \ref{fig:9}). 
At $|G|\sim 1$, Im$S(\omega)$ represents a broad asymmetric peak
centered at $E_{sc}\sim u_0$ with the width $\gamma_{sc}\sim D$, so
that TI-FWM signal exhibits weak oscillations and overall exponential decay
on time scale $|\tau|\sim D^{-1}$. For long time delays, 
$|\tau|\gtrsim D^{-1}$, the
low-frequency singular behavior of the memory functions becomes
important and the crossover to the asymptotic regime 
Eq.\ (\ref{TI-FWM-asympt}) occurs (see curves with $|G|=1.2$ in 
Fig.\ \ref{fig:9}). For larger $|G|$, the crossover
occurs at even longer times, $|\tau|\gg D^{-1}$ (see
curves with $|G|=3.6$ in  Fig.\ \ref{fig:9}).

In Figs.\ \ref{fig:10} and \ref{fig:11} we compare the
numerically calculated TI-FWM signal 
with the analytic solution 
(\ref{Q}-\ref{S}) and with the APM (\ref{S-repul}).
In the latter case,  
the  width of the band of exciton-exciton scattering
states, $\gamma_{sc}$, must be introduced 
phenomenologically
by fitting the memory function $S(\omega)$ with a
Lorentzian [see Figs.\ \ref{fig:10}(b) and \ref{fig:11}(b)].
Even though such an approximation describes 
well the overall lineshape of the memory function,  
the latter deviates from a Lorentzian in the low frequency domain.
These discrepancies have a profound effect on the TI-FWM signal. 
Namely, even though in the initial exponential  regime 
the APM and numerical curves are in a reasonable agreement
with each other, the APM completely  fails 
to reproduce the long nonexponential
tail of the TI-FWM signal. The latter 
originates from the singular low-frequency
behavior of $S(\omega)$ [see Eq.\ (\ref{S-anal1})].
On the other hand, the analytic solution 
(\ref{Q}-\ref{S}) describes accurately the 
long nonexponential tail of the TI-FWM signal
as well as the crossover between the two
regimes 
[see Figs.\ \ref{fig:10}(a) and \ref{fig:11}(a)].

\section{conclusions}
\label{sec:conclusion}

In summary, we studied the role of the exciton-exciton interactions
in the ultrafast nonlinear optical spectroscopy of semiconductor
quantum  wells in a perpendicular magnetic field. 
In the case of attractive exciton-exciton
interactions, we found that the biexcitonic effects dominate.
For repulsive interaction, we have shown that the time-integrated 
four-wave-mixing signal
exhibits a long nonexponential tail for negative time delays. 
We traced the origin of this tail to the low-frequency of the memory
kernel for the two-exciton wave-function, for which we derive an
explicit analytical expression.

Our analytical solution, given by Eqs.\ (\ref{Q}-\ref{S}), was derived
here in the case of strong magnetic field, where the kinetic energy of
the electrons and holes is frozen and the exciton-exciton interactions
play a  dominant role in the third-order optical response. We
believe, however, that Eqs.\ (\ref{Q}-\ref{S}) can serve as an
analytical model also for the zero magnetic field case, for example, when the 
dynamics is dominated by interactions of 1s excitons. In particular,
our solution reproduces, as a limiting case, the well-known average
polarization model,\cite{wegener90,ssr91,schafer96,chemla99,kner99} 
which has been used in the case of zero magnetic field. 
Unlike the APM, however, the model (\ref{Q}-\ref{S})
takes into account nonperturbatively the correlation effects due to the
{\em continuum} of exciton-exciton scattering states.

\acknowledgements
We thank D. S. Chemla for useful discussions.
This work was supported by the NSF grant ECS-9703453.

\appendix
\section{}
\label{app:gen}
In this appendix, we present a general expression for the 
third-order FWM polarization derived using the
canonical-transformation formalism.\cite{perakis99,primozich00,ssr}
The total Hamiltonian of the system is
$H_{tot}=H+H_{p}+H_{s}$, where $H$ is the two-band
semiconductor Hamiltonian (considered in the rotating frame 
from now on) and $H_{\alpha}$ describe the
coupling to the optical fields
($\alpha=1,2$ for probe and pump, respectively),
\begin{equation}
\label{H-opt}
H_{\alpha}=-M_{\alpha}(t)U^{\dag} + {\rm H.c.}
\end{equation}
with 
\begin{equation}
\label{M-U}
M_{\alpha}(t)=e^{i{\bf k}_{\alpha}{\bf r}-i\omega_0t}
\mu{\cal E}_{\alpha}(t),~~~
U^{\dag}=\sum_{\bf k}a_{\bf k}^{\dag}b_{\bf -k}^{\dag}.
\end{equation}
Here $U^{\dag}$ is the optical transition operator, $a_{\bf k}^{\dag}$
and $b_{\bf -k}^{\dag}$ are the creation operators for electron and hole
respectively  with momentum ${\bf k}$; ${\cal E}_{\alpha}(t)$ are the
amplitudes of the optical fields propagating in the probe 
(${\bf k}_1$) and pump (${\bf k}_2$) directions with central frequency
$\omega_0$, and $\mu$ is the dipole matrix element.

The nonlinear optical polarization is given by
$P(t)=\mu\langle \Psi(t)|U|\Psi(t)\rangle$,
where the state $|\Psi(t)\rangle$ satisfies the Schr\"{o}dinger 
equation with Hamiltonian $H_{tot}$.
The third-order  polarization propagating along the FWM direction 
$2 {\bf k}_2  - {\bf k}_1$  can be obtained by expanding the state
$|\Psi(t)\rangle$ to the first order in  the
probe and second order in  the pump field. 
The FWM polarization then takes the form 
$P_{FWM}(t)=e^{i(2{\bf k}_2-{\bf k}_1)\cdot {\bf r}}\tilde{P}(t)$
with\cite{primozich00,perakis99} 
\begin{equation}
\label{pol-fwm}
\tilde{P}(t)=i\mu^2 e^{-i\omega_0t}\int_{-\infty}^{t}dt'{\cal E}_1^{\ast}(t')
\biggl[\langle 0|U e^{- i H (t - t')}U_{FWM}^{\dag}(t')|0\rangle 
- (t\leftrightarrow t')\biggr].
\end{equation}
The above expression has a form similar to the linear polarization 
due to the probe optical field, with the important difference that 
here the single {\em e-h} pair state $U_{FWM}^{\dag}(t)|0\rangle$
depends on the pump optical field,
\begin{equation}
\label{Ufwm}
U_{FWM}^{\dag}|0\rangle = 
UW^{\dag}|0\rangle - {\cal P}^{\dag}U{\cal P}^{\dag}|0\rangle,
\end{equation}
where $|0\rangle$ is the ground state of $H$.
Here the states ${\cal P}^{\dag}(t)|0\rangle$ and
$W^{\dag}(t)|0\rangle$
are the first and second order terms of the expansion 
of the state $|\Phi(t)\rangle$ that evolves from the semiconductor 
ground state under photoexcitation by the {\em pump alone}, 
\begin{equation}
\label{state}
|\Phi(t)\rangle \propto | 0 \rangle 
- e^{i{\bf k}_2\cdot {\bf r}} 
{\cal P}^{\dag}(t) | 0 \rangle 
+ e^{2 i{\bf k}_2\cdot {\bf r}} 
W^{\dag}(t)| 0 \rangle.
\end{equation} 
The {\em single} {\em e-h} pair state ${\cal P}^{\dag}(t)|0\rangle$ 
and the {\em two} {\em e-h} pair state $W^{\dag}(t)|0\rangle$
excited by the pump optical field 
satisfy the equations
\begin{equation}
\label{eqP}
i\partial_t{\cal P}^{\dag}(t)|0\rangle 
=H{\cal P}^{\dag}(t) | 0\rangle 
+\mu {\cal E}_2(t)U^{\dag}|0\rangle,
\end{equation}
and
\begin{equation}
\label{eqW}
i\partial_t W^{\dag}(t)| 0 \rangle=H W^{\dag}(t)|0\rangle
+\mu {\cal E}_2(t)U^{\dag}{\cal P}^{\dag}(t)|0\rangle,
\end{equation}
respectively.\cite{primozich00,perakis99}

The decomposition (\ref{Ufwm}) has a straightforward interpretation.
The first term in the rhs describes the FWM signal coming from
the excitation of two {\em e-h} pairs by the pump followed by 
the de-excitation of one pair by the probe. The second term
describes the excitation of an {\em e-h} pair by the pump, followed by
de-excitation by the probe and  then excitation of an {\em e-h}  pair 
by the pump again. The first term in Eq.\ (\ref{Ufwm}) describes the 
two-exciton contribution to the FWM signal, while the second term
describes the one-exciton contribution.
Note that, in a two-level system, only the second term in
Eq.\ (\ref{Ufwm}) is present. 

\section{}
\label{app:fwm}

Starting from the general formulae in Appendix \ref{app:gen},
we derive here 
a simple expression for the FWM signal coming from the 
exciton-exciton interactions. 
The Pauli blocking contribution is outlined in  Appendix \ref{app:PB}.

Equation (\ref{eqW}) describes the time evolution of the 
two-exciton state from an initial state of 
two {\em non-interacting} excitons
photoexcited by the pump.
It is useful to separate out the contribution 
$W_0^{\dag}(t)| 0 \rangle$ 
due to  non-interacting excitons with zero momentum 
by writing 
$W^{\dag}(t)|0\rangle=W_0^{\dag}(t)|0\rangle+W_{XX}^{\dag}(t)|0\rangle$, 
where the second term 
comes from the exciton-exciton interactions.
In the following we assume that the photoexcited electrons and holes
are confined to their respective LLL's, so that the energy of two
noninteracting magnetoexcitons with zero momentum is simply $2\Omega_0$,  
where $\Omega_0$ is the pump detuning from the zero-momentum exciton
energy (we work in the rotating frame).

The time evolution of the non-interacting 
exciton state is determined by the equation 
\begin{equation} 
\label{eqW0}
i\partial_tW_0^{\dag}(t)| 0 \rangle =
2 \Omega_0 W_0^{\dag}(t)| 0 \rangle  
+\mu {\cal E}_2(t)U^{\dag}{\cal P}^{\dag}(t)| 0 \rangle.
\end{equation}
After subtracting Eq.\ (\ref{eqW0}) from Eq.\ (\ref{eqW}),
we obtain that the  exciton-exciton interactions 
are  described by 
the time-dependent Schr\"{o}dinger-like equation
\begin{equation} 
\label{eqWXX}
i\partial_tW_{XX}^{\dag}(t)| 0 \rangle =
HW_{XX}^{\dag}(t)| 0 \rangle  
+ (H-2 \Omega_0 ) W_0^{\dag}(t)| 0 \rangle.
\end{equation}
%
The source term on the rhs is the  Hartree-Fock interaction between two 
zero-momentum excitons. The  exciton-exciton correlations come from
the Coulomb potential in the Hamiltonian $H$  
(first term in the rhs of Eq.\ \ (\ref{eqWXX})).

Thus, we can present
Eq.\ (\ref{Ufwm}) as a sum of two contributions:
\begin{equation}
\label{Ufwm1}
U_{FWM}^{\dag}|0\rangle =
U_{PB}^{\dag}|0\rangle+U_{XX}^{\dag}|0\rangle,
\end{equation}
where
\begin{equation}
\label{UPB}
U_{PB}^{\dag}(t)|0\rangle = 
UW_0^{\dag}|0\rangle-{\cal P}^{\dag}U{\cal P}^{\dag}|0\rangle
\end{equation}
describes the Pauli blocking effects for noninteracting
excitons, and
\begin{equation}
\label{Uxx}
U_{XX}^{\dag}(t)|0\rangle = UW_{XX}^{\dag}|0\rangle
\end{equation}
comes from the exciton-exciton interactions. 
The polarization (\ref{pol-fwm}) can then be presented
as a sum of
Pauli blocking and exciton-exciton interaction parts,
$\tilde{P}(t)=\tilde{P}_{PB}(t)+\tilde{P}_{XX}(t)$, where
\begin{eqnarray}
\label{pol-fwm1}
\tilde{P}_{PB}(t)=i\mu^2 e^{-i\omega_0t}\int_{-\infty}^{t}dt'
{\cal E}_1^{\ast}(t')
\biggl[\phi_{PB}(t,t')-\phi_{PB}(t',t)\biggr],
\nonumber\\
\tilde{P}_{XX}(t)=i\mu^2 e^{-i\omega_0t}\int_{-\infty}^{t}dt'
{\cal E}_1^{\ast}(t')
\biggl[\phi_{XX}(t,t')-\phi_{XX}(t',t)\biggr],
\end{eqnarray}
with 
\begin{eqnarray}
\label{phi}
\phi_{PB}(t,t')=\langle 0|U e^{- i H (t - t')}U_{PB}^{\dag}(t')|0\rangle ,
\nonumber\\
\phi_{XX}(t,t')=\langle 0|U e^{- i H (t - t')}U_{XX}^{\dag}(t')|0\rangle.
\end{eqnarray}
Equations (\ref{pol-fwm1}) provide an exact  formal
expression for the third-order FWM signal.

Here we focus on the 
the interaction-induced FWM polarization, 
and defer the Pauli blocking contribution to Appendix \ref{app:PB}.
$\tilde{P}_{XX}(t)$ 
is determined by the correlation function
$\phi_{XX}(t,t')$, which satisfies the
equation
\begin{equation} 
\label{eqphiXX}
i\partial_t\phi_{XX}(t,t')=
\langle 0|UHe^{-iH(t-t')}U_{XX}^{\dag}(t')|0\rangle
\end{equation}
with the initial condition $\phi_{XX}(t,t)=\chi(t)$, where
\begin{equation} 
\label{chi}
\chi(t)=
\langle 0|UU_{XX}^{\dag}(t)|0\rangle
=\langle 0|UUW_{XX}^{\dag}(t)|0\rangle.
\end{equation}
Since the state $U^{\dag}|0\rangle=N^{1/2}d_0^{\dag}|0\rangle$ is an
eigenstate of $H$ [see Eq.\ (\ref{HX})],
we obtain from Eq.\ (\ref{eqphiXX}) that
\begin{equation} 
\label{phiXX}
\phi_{XX}(t,t')=e^{-i\Omega_0(t-t')-\Gamma |t-t'|}\chi(t'),
\end{equation}
where $\Gamma$ is
the {\em exciton} homogeneous broadening. 
Since   Eq.\ (\ref{phiXX}) is linear in the probe 
optical field,  we consider for simplicity a 
$\delta$-function probe   
${\cal E}_1(t)=e^{-i\omega_0\tau}{\cal E}_1\delta(t+\tau)$
Substituting this into Eq.\ (\ref{phi}) 
we obtain 
\begin{eqnarray}
\label{fwm-final2}
\tilde{P}_{XX}(t,\tau)=i\mu^2{\cal E}_1 \theta(t+\tau) 
e^{- \Gamma ( t + \tau)-i \omega_0(t-\tau)}
\biggl[e^{-i\Omega_0(t+\tau)}\chi(-\tau) 
- e^{i\Omega_0(t+\tau)}\chi(t) \biggr].
\end{eqnarray}

Using Eq.\ (\ref{expan}), $\chi(t)$ can be expressed 
in terms of the  coefficients of the expansion of
the state $W_{XX}^{\dag}(t)|0\rangle$
in the two-exciton basis (\ref{Psi-twoexc}),
\begin{equation} 
\label{Wp}
w_p(t)=\langle {\bf p, -p}|W_{XX}^{\dag}(t)|0\rangle,
\end{equation}
as follows: 
\begin{equation} 
\label{chi1}
\chi(t)=w_0-\frac{1}{N}\sum_{\bf p} w_p=w_0-\tilde{w}_0.
\end{equation}
The two terms in the rhs of Eq.\ (\ref{chi1}) reflect the two possible
ways of arranging four particles into two excitons.
The equation for the amplitude 
$w_p(t)$ follows from Eq.\ (\ref{eqWXX}),
\begin{equation} 
\label{eqWXX1}
i\partial_tw_p(t)=
\langle {\bf p, -p}|HW_{XX}^{\dag}(t)| 0 \rangle  
+ \langle {\bf p, -p}|(H - 2\Omega_0 ) W_0^{\dag}(t)|0\rangle.
\end{equation}
After projecting out the rhs of the above equation
{\bf }in the two-exciton basis Eq.\ (\ref{Psi-twoexc}) 
and using Eq.\ (\ref{H-twoexc}) for the matrix elements of $H$,
Eq.\ (\ref{eqWXX1}) takes the form
\begin{equation} 
\label{eqWXX2}
i\partial_t w_p(t)-2\Omega_pw_p(t)=
-\int\frac{d{\bf q}}{(2\pi)^2}V({\bf p},{\bf q})w_q(t)
+2(\epsilon_p-\epsilon_0)w^0_p(t)
-\int\frac{d{\bf q}}{(2\pi)^2}V({\bf p},{\bf q})w^0_q(t),
\end{equation}
where $w^0_p(t)=\langle {\bf p, -p}|W_0^{\dag}(t)|0\rangle$
is the noninteracting two-exciton amplitude.
The equation for the latter
can be obtained straightforwardly from Eq.\ (\ref{eqW0}):
\begin{equation} 
\label{eqW01}
i\partial_t w^0_p(t)-2(\Omega_0-i\Gamma)w^0_p(t)=
\mu {\cal E}_2(t)\langle {\bf p, -p}|U^{\dag}{\cal P}^{\dag}(t)|0\rangle.
\end{equation}
In order to evaluate the last term in Eq.\ (\ref{eqW01}), we use 
Eq.\ (\ref{eqP}) for the one-exciton state 
${\cal P}^{\dag}(t)|0\rangle$ to obtain
\begin{equation} 
(i\partial_t -\Omega_0+i\Gamma)
\langle {\bf p, -p}|U^{\dag}{\cal P}^{\dag}(t)|0\rangle
=\mu {\cal E}_2(t)\langle {\bf p, -p}|U^{\dag}U^{\dag} |0\rangle.
\end{equation}
Using Eqs. (\ref{useful}) and (\ref{expan}), we find that 
$\langle {\bf p, -p}|U^{\dag}U^{\dag} |0\rangle=N\delta_{{\bf p}0}-1$,
so that
\begin{equation} 
\label{RHS}
\langle {\bf p, -p}|U^{\dag}{\cal P}^{\dag}(t)|0\rangle=
(N\delta_{{\bf p}0}-1)p(t)=\Biggl[\frac{2 \pi}{l^2}
\,\delta({\bf p})-1\Biggr]p(t),
\end{equation}
where $p(t)$ is the linear pump-induced polarization, satisfying
\begin{equation}
\label{eqp1}
i\partial_t p(t)=(\Omega_0-i\Gamma)p(t)+\mu{\cal E}_2(t).
\end{equation}
Substituting Eq.\ (\ref{RHS}) into Eq.\ (\ref{eqW01}), and comparing the
latter to Eq.\ (\ref{eqp1}), we obtain that 
\begin{equation} 
\label{W02}
w^0_p(t)=\Biggl[\frac{2 \pi}{l^2}\, \delta({\bf p})-1\Biggr]\frac{p^2(t)}{2},
\end{equation}
Substituting the above expression for 
$w^0_p(t)$ into Eq.\ (\ref{eqWXX2}) and using
the decomposition 
$V({\bf p},{\bf q})=V_S({\bf p},{\bf q})+V_A({\bf p},{\bf q})$
together with the relation (\ref{ident}),
we arrive at the following equation 
for the two-exciton amplitude:
\begin{eqnarray} 
\label{eqWXX3}
i\partial_t w_p(t)-2\Omega_pw_p(t)=
-\int\frac{d{\bf q}}{(2\pi)^2}V({\bf p},{\bf q})w_q(t)
+V_p^{HF}p^2(t),
\end{eqnarray}
where
\begin{eqnarray} 
\label{hartree}
V_p^{HF}=\frac{1}{2}\int\frac{d{\bf q}}{(2\pi)^2}V_A({\bf p},{\bf q})
-\frac{V_A({\bf p},0)}{4\pi l^2},
\end{eqnarray}
According to Eq.\ (\ref{chi1}), the amplitude $\chi(t)$ is determined
by the ${\bf p}=0$ solution of Eq.\ (\ref{eqWXX3}) and by 
$\tilde{w}_0(t)=N^{-1}\sum_{\bf p}w_p(t)$. The equation for $\tilde{w}_0(t)$
can be straightforwardly obtained by summing  
Eq.\ (\ref{eqWXX3}) 
over all momenta ${\bf p}$ and using the relation (\ref{ident}). This gives
\begin{eqnarray} 
\label{eqTWXX}
i\partial_t \tilde{w}_0(t)-2\Omega_0\tilde{w}_0(t)=
-\frac{1}{N}\sum_{\bf p}\int\frac{d{\bf q}}{(2\pi)^2}V_A({\bf p},{\bf q})w_q(t)
+\frac{1}{N}\sum_{\bf p}V_p^{HF}p^2(t).
\end{eqnarray}
Setting ${\bf p}=0$ in Eq. (\ref{eqWXX3}), and subtracting from the
latter  Eq. (\ref{eqTWXX}), we obtain
a simple equation for the amplitude 
$\chi(t)$ (Eq.\ (\ref{chi1})) that determines the FWM polarization,
\begin{eqnarray} 
\label{eq-chi}
i\partial_t \chi(t)-2\Omega_0\chi(t)=
V^{HF}p^2(t) + F(t),
\end{eqnarray}
where 
\begin{eqnarray} 
\label{hartree1}
V^{HF}=V_0^{HF}-\frac{l^2}{2 \pi} \int d{\bf q} \, V_q^{HF}
\end{eqnarray}
is the Hartree-Fock interaction, and 
\begin{eqnarray} 
\label{corr}
F(t)=\frac{l^2}{\pi} \int d{\bf q} \, V_q^{HF} \, w_q(t)
\end{eqnarray}
describes the effects of the exciton-exciton correlations.

\section{}
\label{app:PB}
In this Appendix we derive the Pauli blocking contribution to the
FWM signal in strong magnetic field, determined by Eq.\  (\ref{UPB}).
Using  Eq.\  (\ref{eqP}), the second term in Eq.\  (\ref{UPB})
satisfies
\begin{equation} 
i \partial_t {\cal P}^{\dag}(t) U {\cal P}^{\dag}(t)| 0 \rangle
= 2(\Omega_0 - i \Gamma)
{\cal P}^{\dag}(t) U {\cal P}^{\dag}(t)| 0 \rangle
+\mu {\cal E}_2(t)[U^{\dag} U {\cal P}^{\dag}(t)
+{\cal P}^{\dag}(t) U U^{\dag} ]| 0 \rangle
\end{equation} 
where we used the fact that 
${\cal P}^{\dag}(t)| 0 \rangle$ is the zero-momentum 
exciton state with energy $\Omega_0$. 
Using the relation $U^{\dag} = N^{1/2} d^{\dag}_{0}$
[see Eq.\ (\ref{d})]
and noting that $d^{\dag}_{0} d_0=1$ 
when acting on the one-exciton state  
${\cal P}^{\dag}(t)| 0 \rangle$, 
while 
$U U^{\dag} | 0 \rangle = 
N d_{0} d^{\dag}_{0} | 0\rangle=  N | 0\rangle$, 
we obtain that
\begin{equation} 
\label{psf1} 
i \partial_t {\cal P}^{\dag}(t) U {\cal P}^{\dag}(t)| 0 \rangle
= \Omega_0 {\cal P}^{\dag}(t) U {\cal P}^{\dag}(t)| 0 \rangle
+ 2N \mu {\cal E}_2(t){\cal P}^{\dag}(t)| 0 \rangle.
\end{equation} 
The non-interacting contribution to the 
two-exciton state, $W^{\dag}_0 | 0 \rangle$, 
is determined by Eq.\ (\ref{eqW0}).
After projecting with $U$ on the lhs
we obtain  that 
\begin{equation} 
\label{EqW0}
i\partial_t UW_0^{\dag}(t)| 0 \rangle =
2\Omega_0 UW_0^{\dag}(t)| 0 \rangle  
+\mu {\cal E}_2(t)U U^{\dag}{\cal P}^{\dag}(t)| 0 \rangle.
\end{equation}
The last term can be simplified by using 
$U U^{\dag}{\cal P}^{\dag}(t)| 0 \rangle
=Nd_0 d_0^{\dag}{\cal P}^{\dag}(t)| 0 \rangle
=2(N-1){\cal P}^{\dag}(t)| 0 \rangle$, so that 
\begin{equation} 
i\partial_t UW_0^{\dag}(t)| 0 \rangle =
\Omega_0 UW_0^{\dag}(t)| 0 \rangle  
+2(N-1) \mu {\cal E}_p(t){\cal P}^{\dag}(t)| 0 \rangle.
\end{equation}
Comparing the above  equation
to (\ref{psf1}),
we see that $UW_0^{\dag}(t)|0 \rangle 
= (1-1/N) {\cal P}^{\dag}(t) U {\cal P}^{\dag}(t)| 0 \rangle$, and
we thus obtain 
\begin{equation} 
\label{PB}
U_{PB}^{\dag}(t)|0\rangle = -\frac{1}{N} {\cal P}^{\dag}(t)
 U {\cal P}^{\dag}(t)| 0 \rangle.
\end{equation} 
Using the formal solution of  Eq.\ (\ref{eqP}),
\begin{equation} 
{\cal P}^{\dag}(t)
= -i \mu \int_{-\infty}^{t} 
dt' {\cal E}_2(t') \ e^{-i H(t-t')} U^{\dag} e^{i H (t-t')},
\end{equation}
and that ${\cal P}^{\dag}(t) | 0 \rangle 
= p(t) U^{\dag} | 0 \rangle$, where the pump polarization 
$p(t)$ is given by Eq.\ (\ref{p}), 
we obtain 
\begin{equation} 
\phi_{PB}(t,t')=
 \frac{i \mu}{N} p(t') \int_{-\infty}^{t'} dt'' 
{\cal E}_2(t'') 
\langle 0|U e^{- i H (t - t')}
\ e^{-i H(t'-t'')} U^{\dag} e^{i H (t'-t'')}
UU^{\dag} | 0 \rangle.
\end{equation} 
Finally, using 
$UU^{\dag} | 0 \rangle = N | 0\rangle$
and  that $U^{\dag} | 0\rangle$
is proportional to the magnetoexciton eigenstate of $H$,
with eigenvalue $\Omega_0$, 
we obtain that 
\begin{equation} 
\phi_{PB}(t,t')=
 i \mu  p(t') \int_{-\infty}^{t'} dt'' 
{\cal E}_2(t')  e^{- i \Omega_0 (t - t'')
- \Gamma |t - t''|}.
\end{equation} 
The Pauli blocking contribution to the FWM polarization 
is then  calculated by substituting
the above expression into Eq.\ (\ref{pol-fwm1}). 

\section{} 
\label{app:APM}

In this Appendix we show the connection between our
expression for the FWM signal,
Eqs.\ (\ref{pol-fwm1}), 
(\ref{phi}), and (\ref{phiXX}), and that of 
Refs. \onlinecite{chemla99,kner99,schafer96}. 
The equation for the four-wave-mixing polarization 
$P_{XX}(t)=e^{i\omega_0t}\tilde{P}_{XX}(t)$
can be obtained by taking the time  derivative with respect to 
$t$ of the rhs of Eq.\ (\ref{pol-fwm1}),
\begin{equation} 
\label{eom}
i \partial_t P_{XX}(t)=
i\mu^2 \int_{-\infty}^{t}dt'
{\cal E}_1^{\ast}(t')
\biggl[i \partial_t \phi_{XX}(t,t')-
i \partial_t \phi_{XX}(t',t)\biggr].
\end{equation}
From Eq.\ (\ref{phiXX}) 
we obtain after some algebra that 
\begin{equation}
\label{deriv1}
i \partial_t \phi_{XX}(t,t') =
( \Omega_0 - i \Gamma)  \phi_{XX}(t,t') 
\end{equation} 
and 
\begin{equation}
\label{deriv2} 
i \partial_t \phi_{XX}(t',t) =
( \Omega_0 - i \Gamma)  \phi_{XX}(t',t) 
+ \biggl[i \partial_t \chi(t)-
2 \Omega_0 \chi(t) \biggr]
e^{i \Omega_0 (t - t')- \Gamma (t - t')}.
\end{equation}
Substituting  Eqs.\ (\ref{deriv1}) and  (\ref{deriv2})
into Eq.\ (\ref{eom}) and  using  Eq.\ (\ref{eq-chi}) 
we obtain a familiar  equation for 
the FWM polarization: 
\begin{equation} 
\label{eom1}
i \partial_t P_{XX}(t) =
( \Omega_0 - i \Gamma) P_{XX}(t)
- \mu p_{1}^{*}(t) 
\biggl[V^{HF} p^{2}(t) + F(t) \biggr],
\end{equation}
where $p_{1}(t)$ is 
the probe-induced polarization
\begin{eqnarray} 
p_{1}(t)=-i\mu\int_{-\infty}^{t}dt'
e^{-i(\Omega_0-i\Gamma)(t-t')}{\cal E}_1(t').
\end{eqnarray}
The simple (momentum-independent) form of the above equation is due
to the dispersionless energy spectrum of
the  electron and hole in the magnetic
field. At zero field, a similar equation is obtained after averaging
out over the spatial degrees of
freedom.\cite{schafer96,chemla99,kner99}
The function $F(t)$, given by Eq.\ (\ref{corr}), is determined by the
two-exciton amplitude $w_p(t)$, satisfying Eq.\ (\ref{eqWXX1}), and
describes memory effects due to four-particle correlations. The 
APM\cite{wegener90,ssr91,schafer96,chemla99,kner99} is recovered if one
uses the simple model (\ref{GAPM}) for the memory function $S(\omega)$.

\section{} 
\label{app:tail}

In this Appendix, we derive the long-time asymptotics of
$\chi(t)$. For delta-function pump centered at $t=0$, the pump
polarization is simply 
$p(t)=-i\mu {\cal E}_2\theta(t)e^{-i\Omega_0 t-\Gamma t}$, so that
the Fourier transform of $p^2(t)$ is
\begin{equation}
\label{P2-fourier}
p^{(2)}(\omega)=\frac{-i(\mu {\cal E}_2)^2}{\omega-2\Omega_0+2i\Gamma}.
\end{equation} 
Note that for short pump duration, we can write 
$\chi(t)=e^{-i2\Omega_0t}\tilde{\chi}(t)$, where $\tilde{\chi}(t)$ is
the amplitude for zero detuning; therefore, it is
sufficient to evaluate the Fourier transform of Eq.\ (\ref{chi-anal})
at $\Omega_0=0$. 
For $t\gg \Gamma^{-1}$, the relevant frequencies are small,
$\omega\ll\Gamma$, so that  
$p^{(2)}(\omega)\simeq -(\mu {\cal E}_2)^2/2\Gamma$, and 
$\tilde{\chi}(t)$ takes the form

\begin{equation}
\label{chi-tilde}
\tilde{\chi}(t)=
-\frac{(\mu {\cal E}_2)^2}{4\pi\Gamma}V^{HF}
\int_{-\infty}^{\infty}\frac{d\omega}{\omega}e^{-i\omega t}S(\omega).
\end{equation}
In order to get $\tilde{\chi}(t)$ for long times,
$t\gg D^{-1}$, we substitute the low-frequency asymptotic expression 
(\ref{S-anal1}) for the memory function (with $\Omega_0=0$), 
\begin{equation}
\label{chi-tilde1}
\tilde{\chi}(t)=
-\frac{(\mu {\cal E}_2)^2}{4\pi\Gamma}V^{HF}[J_1(t)+J_2(t)],
\end{equation}
where we separated out the contributions from the real and imaginary parts
of $S(\omega)$:
\begin{equation}
\label{J1}
J_1(t)=
\int_{-\infty}^{\infty}\frac{d\omega}{\omega}
\frac{e^{-i\omega t}\Bigl(1+|G|\ln|\frac{2D}{\omega}|\Bigr)}
{\Bigl(1+|G|\ln|\frac{2D}{\omega}|\Bigr)^2+\pi^2 |G|^2\theta(\omega)}
\end{equation}

\begin{equation}
\label{J2}
J_2(t)=-i
\int_{0}^{\infty}\frac{d\omega}{\omega}
\frac{e^{-i\omega t}\pi |G|}
{\Bigl(1+|G|\ln|\frac{2D}{\omega}|\Bigr)^2+\pi^2 |G|^2\theta(\omega)}.
\end{equation}
Consider first $J_1(t)$. For $t\gg D^{-1}$, the characteristic
frequencies are small, so that the second term in the denominator of
the integrand can be neglected as compared to the logarithmic
term. In this case, the main contribution comes from the imaginary part
of the integrand,
\begin{equation}
\label{J1-final}
J_1(t)\simeq 
-i\int_{-\infty}^{\infty}\frac{d\omega}{\omega}
\frac{\sin\omega t}{1+|G|\ln|\frac{2D}{\omega}|}
\simeq \frac{-i\pi}{1+|G|\ln 2Dt}.
\end{equation}
Turning to $J_2(t)$, we notice that, for $t\gg D^{-1}$, the main
contribution comes from the real part of the integrand. After
omitting the second term in the denominator and integrating by parts,
we obtain
\begin{equation}
\label{J2-final}
J_2(t)\simeq 
-i\pi t\int_{0}^{\infty}\frac{d\omega \sin\omega t}
{1+|G|\ln|\frac{2D}{\omega}|}
\simeq \frac{-i\pi}{1+|G|\ln 2Dt}.
\end{equation}
One can show that the subleading terms are
$\sim |G|\Bigl(1+|G|\ln 2Dt\Bigr)^{-2}$. Note that the imaginary and real
parts of $S(\omega)$ contribute  {\em equally} to the long-time
asymptotics. Finally, substituting Eqs.\ (\ref{J1-final}) and
(\ref{J2-final}) into Eq.\ (\ref{chi-tilde1}), 
we arrive at Eq.\ (\ref{chi-asympt}). 



\begin{figure}
\caption{
The memory function $S(\omega)$ for
attractive exciton-exciton interaction calculated from
Eq. (\ref{S}) with $a/l=3.0$ and $u_0/E_0=0.05$ (solid line),
$u_0/E_0=0.4$ (dotted line), and $u_0/E_0=0.8$ (dashed line). The
oscillator strength of the biexciton resonance 
increases with $u_0$.
}
\label{fig:1}
\end{figure}

\begin{figure}
\caption{
The memory function $S(\omega)$ for
attractive exciton-exciton interaction with $a/l=3.0$ and
$u_0/E_0=-0.05$ (solid line), $u_0/E_0=-0.4$ (dotted line), and $u_0/E_0=-0.8$
(dashed line). With increasing $|u_0|$, the asymmetric band of
exciton-exciton scattering states transforms into a narrow antibound
resonance. 
}
\label{fig:2}
\end{figure}

\begin{figure}
\caption{
Calculated TI-FWM signal for short pump duration, $t_0 E_0=0.5$, for
$a/l=6.0$, $\Gamma/E_0=0.0125$, $\gamma_b/E_0=0.005$, $\Omega_0=0$, with
$u_0/E_0=0.05$ (solid curve) and $u_0/E_0=0.4$ (dashed curve). 
The period of the biexciton oscillations is
$T_b=2\pi/E_b\simeq 2\pi/u_0$.
}
\label{fig:3}
\end{figure}

\begin{figure}
\caption{
Calculated SR-FWM signal for 
$u_0/E_0=0.4$, and the rest of the parameters same as in 
Fig.\ \ref{fig:3}. The exciton peak strength oscillates with the
biexciton period $T_b=2\pi/E_b\simeq 2\pi/u_0$.
}
\label{fig:4}
\end{figure}

\begin{figure}
\caption{
Calculated TI-FWM signal at intermediate magnetic field, $a/l=3.0$,
for $u_0/E_0=0.05$ (solid line),
$u_0/E_0=0.4$ (dotted line), and $u_0/E_0=0.8$ (dashed line)
with (a) short pump duration,
$t_0 E_0=0.5$, and (b) long  pump duration,
$t_0 E_0=5.0$. The rest of parameters same as in Fig.\ \ref{fig:3}.
}
\label{fig:5}
\end{figure}

\begin{figure}
\caption{
Calculated SR-FWM signal for $u_0=0.4$, pump tuned to exciton,
$\Omega_0=0$, and long pulse duration, $t_0 E_0=5.0$. 
The biexciton peak is 
suppressed while the exciton peak decays with characteristic time
$(2\Gamma)^{-1}$ for $\tau>0$. The rest of parameters
same as in Fig.\ \ref{fig:5}.
}
\label{fig:6}
\end{figure}

\begin{figure}
\caption{
Calculated SR-FWM signal for the pump tuned to biexciton, $2\Omega_0=E_b$.
For $\tau>0$, biexciton peak decays with characteristic time
$(2\Gamma)^{-1}$ while exciton peak follows the pump. The rest of parameters
same as in Fig.\ \ref{fig:6}.
}
\label{fig:7}
\end{figure}

\begin{figure}
\caption{
Comparison of TI-FWM signals calculated using the analytical solution
(\ref{Q}-\ref{S}) (dotted curve) and the APM (\ref{GAPM}) (dashed
curve) with the exact numerical calculations (solid curve) for
$a/l=1.5$,  $u_0/|E_0|=0.4$ with (a) $\gamma_b=0.005E_0$ and (b) 
$\gamma_b=2\Gamma=0.025E_0$. The rest of parameters
same as in Fig.\ \ref{fig:3}.
}
\label{fig:8}
\end{figure}

\begin{figure}
\caption{
Calculated TI-FWM signal for repulsive interaction with $G=-6.0$
(solid line), $G=-4.8$ (dotted line), $G=-3.6$ (dashed line), 
$G=-1.2$ (long-dashed line), and $G=-0.24$ (dot-dashed line),
with (a) $\Gamma/E_0=0.0125$ and (b) $\Gamma/E_0=0.05$.
For larger $|G|$, the decay is exponential with the
interaction-induced characteristic time $\gamma_{sc}^{-1}$.
For smaller $|G|$, the exponential decay is followed by the
nonexponential tail for large $|\tau|$. 
The rest of the parameters are the same as in Fig.\ \ref{fig:3}.
}
\label{fig:9}
\end{figure}

\begin{figure}
\caption{
(a) Comparison of TI-FWM signals calculated
using analytical solution (\ref{Q}-\ref{S}) (dotted curve) and the APM
(\ref{S-repul}) (dashed curve) with the exact numerical calculations
(solid curve) for 
$a/l=3.0$,  $G=-1.2$ with $\Gamma/E_0$ and $t_0E_0=0.5$. The
exponential decay at initial time delays is followed by the
power law decay for large $|\tau|$. 
(b) Real (upper curve) and imaginary (lower curve) parts of the
memory function fitted by a Lorentzian (dotted line). The Lorentzian
width determines the exponential decay of the signal in the APM. The
deviation of $S(\omega)$ from the Lorentzian at low frequencies
leads to the power-law decay for large $|\tau|$.
}
\label{fig:10}
\end{figure}

\begin{figure}
\caption{Same as Fig.\ \ref{fig:10}, but for weaker magnetic field, 
$a/l=1.5$. The crossover to the nonexponential regime occurs at earlier
time delays. 
}
\label{fig:11}
\end{figure}


\clearpage
\begin{center}
\epsfxsize=5.0in
\epsffile{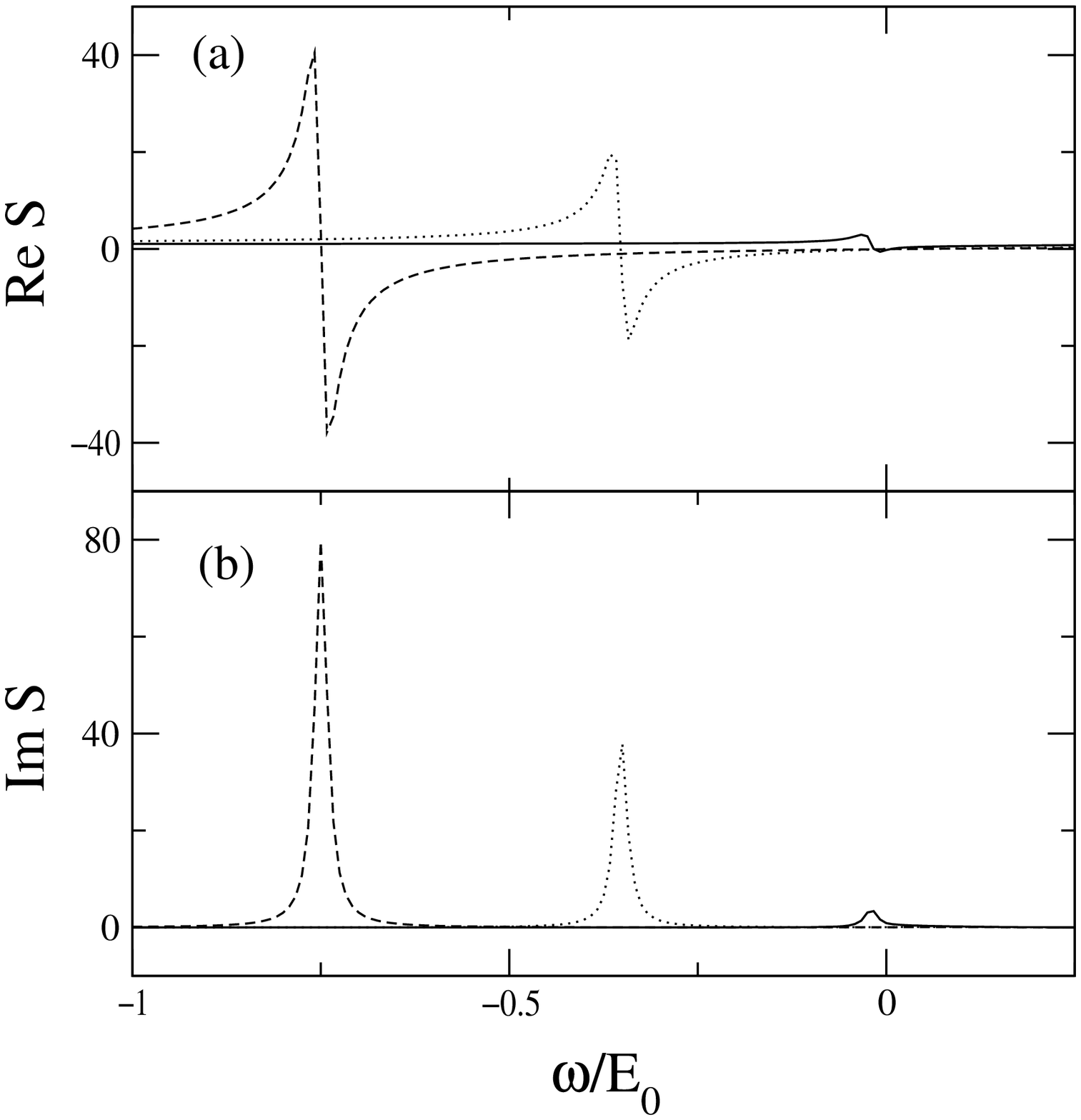}
\end{center}
\vspace{40mm}
\centerline{Fig. 1}

\clearpage
\begin{center}
\epsfxsize=5.0in
\epsffile{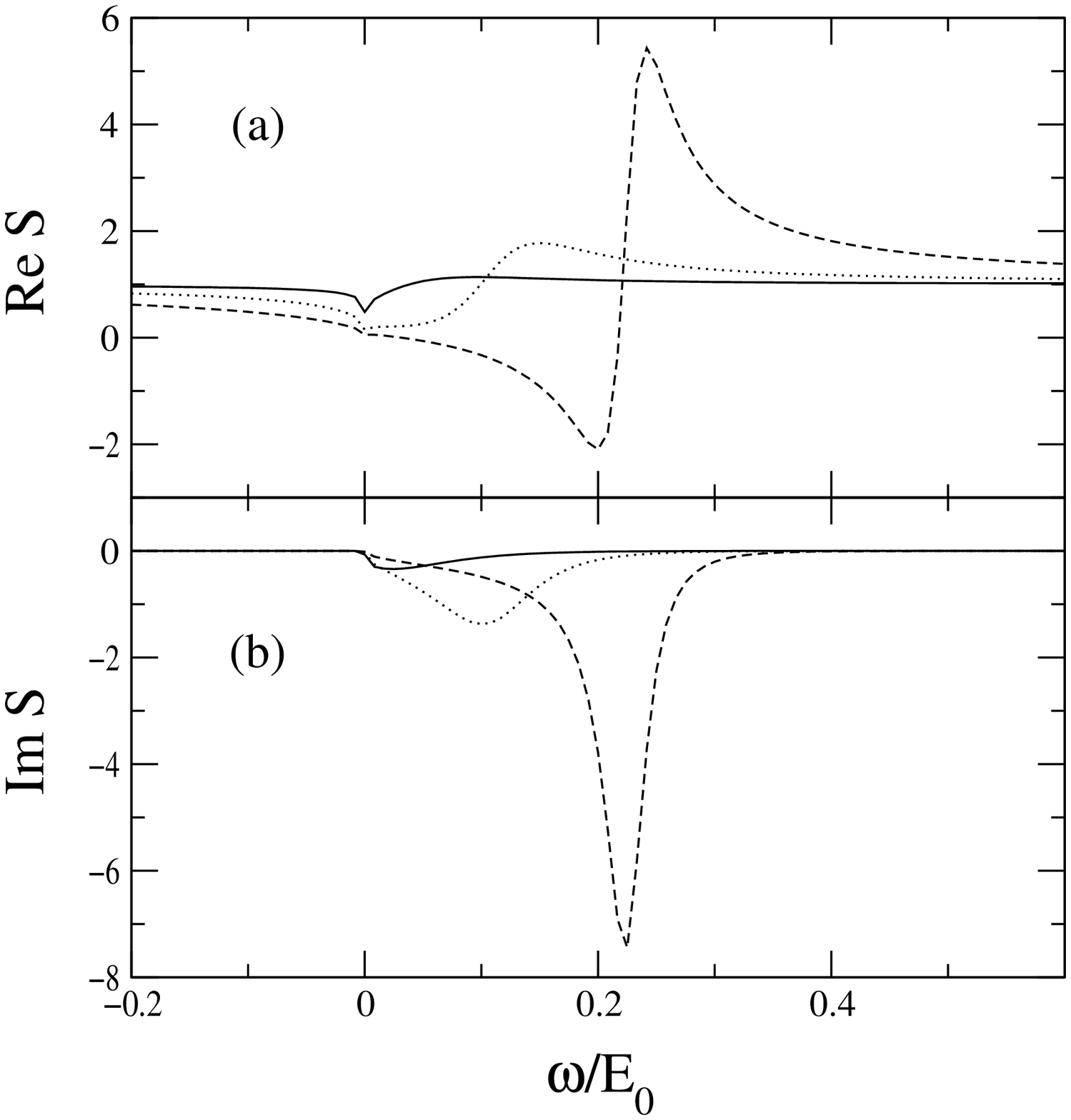}
\end{center}
\vspace{40mm}
\centerline{Fig. 2}
\clearpage
\begin{center}
\epsfxsize=5.0in
\epsffile{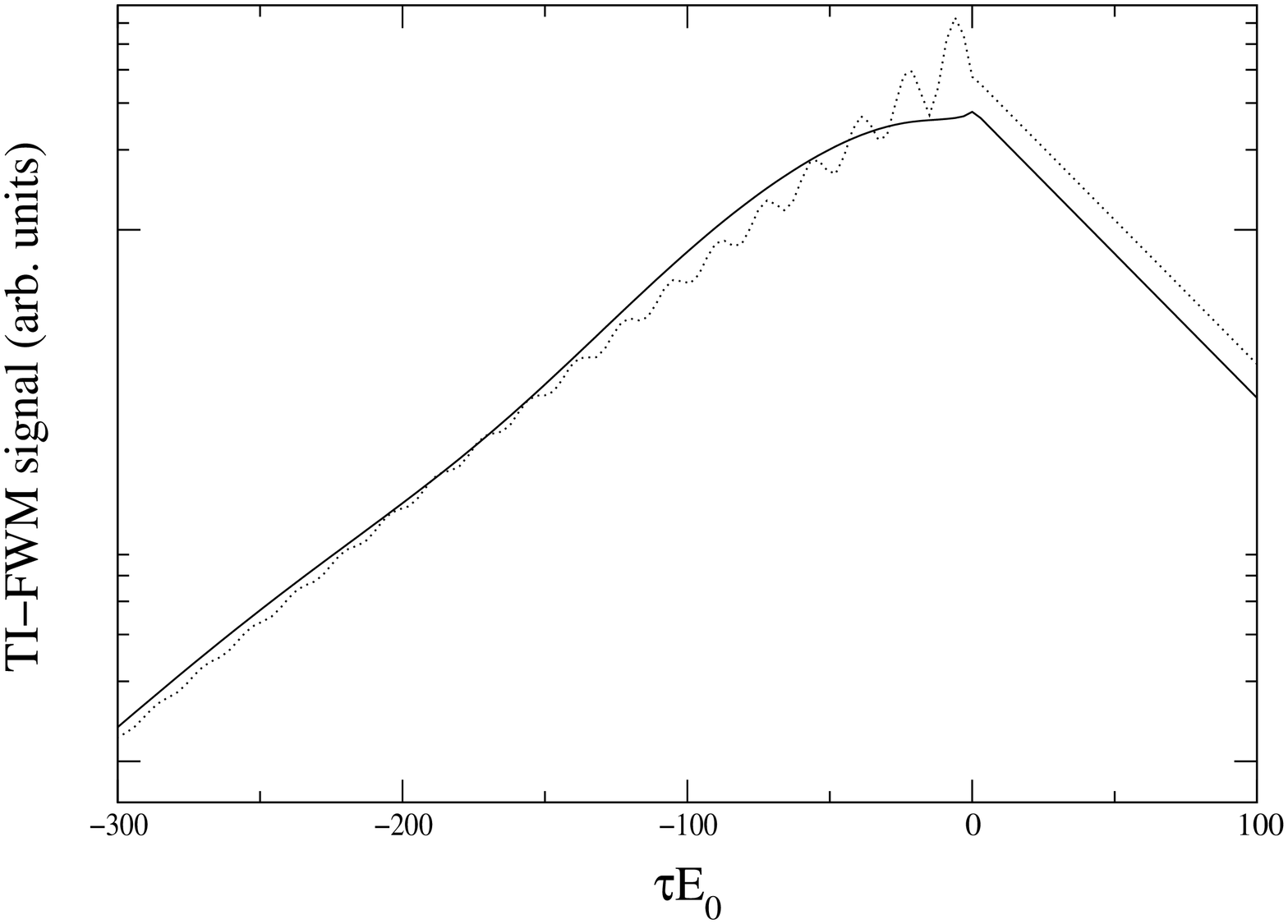}
\end{center}
\vspace{80mm}
\centerline{Fig. 3}
\clearpage
\begin{center}
\epsfxsize=6.0in
\epsffile{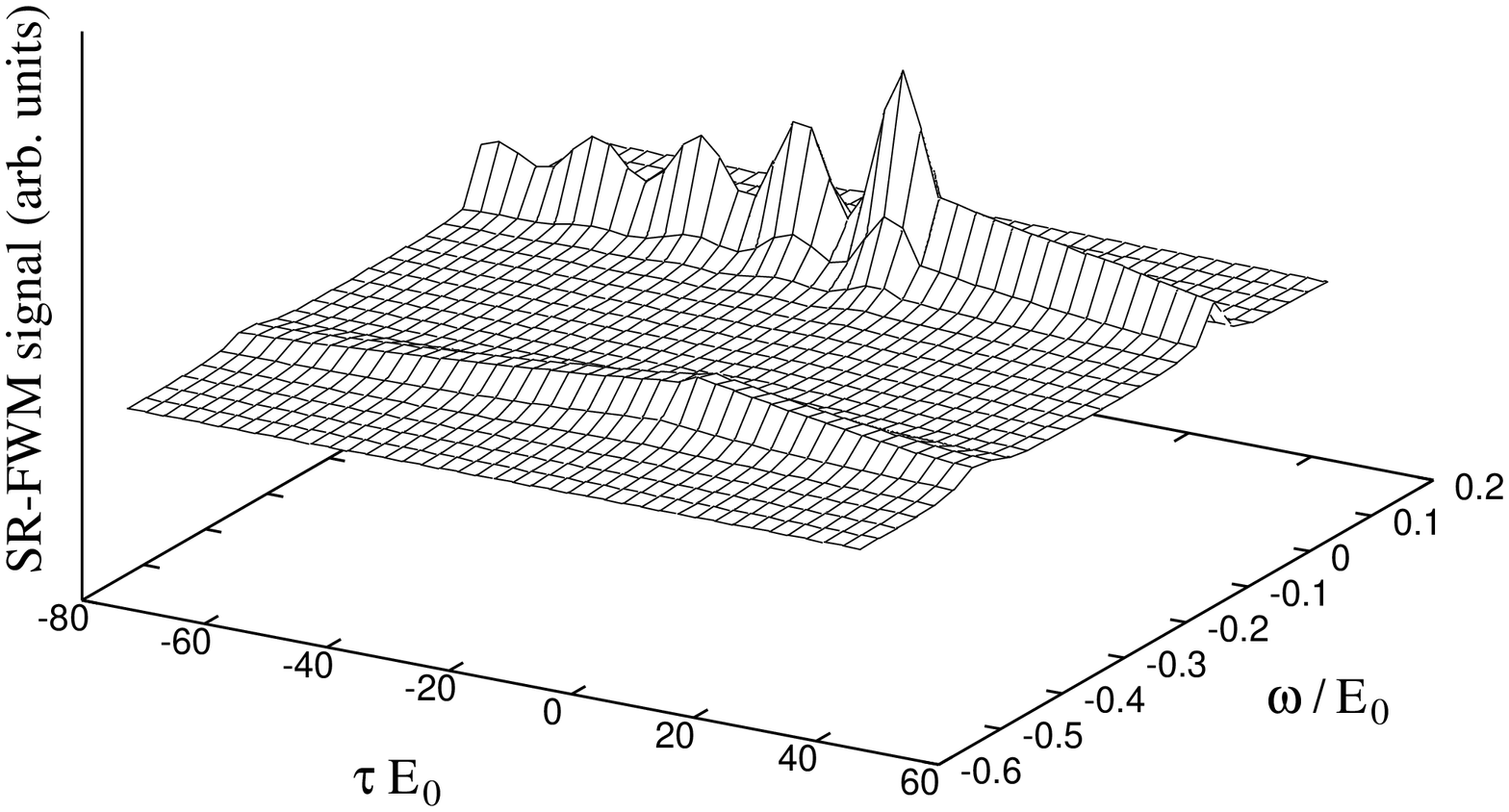}
\end{center}
\vspace{80mm}
\centerline{Fig. 4}
\clearpage
\begin{center}
\epsfxsize=5.0in
\epsffile{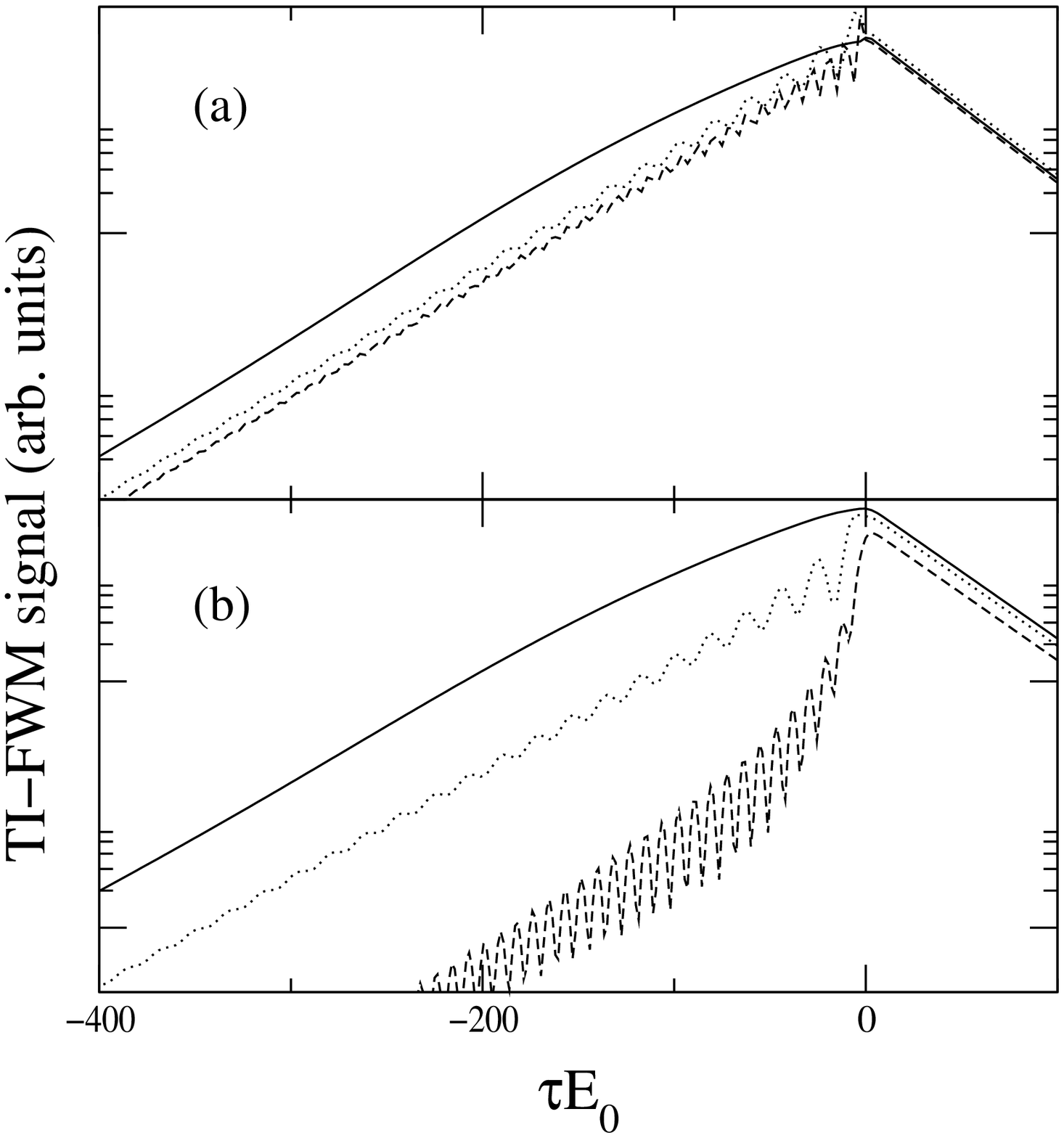}
\end{center}
\vspace{40mm}
\centerline{Fig. 5}
\clearpage
\begin{center}
\epsfxsize=6.0in
\epsffile{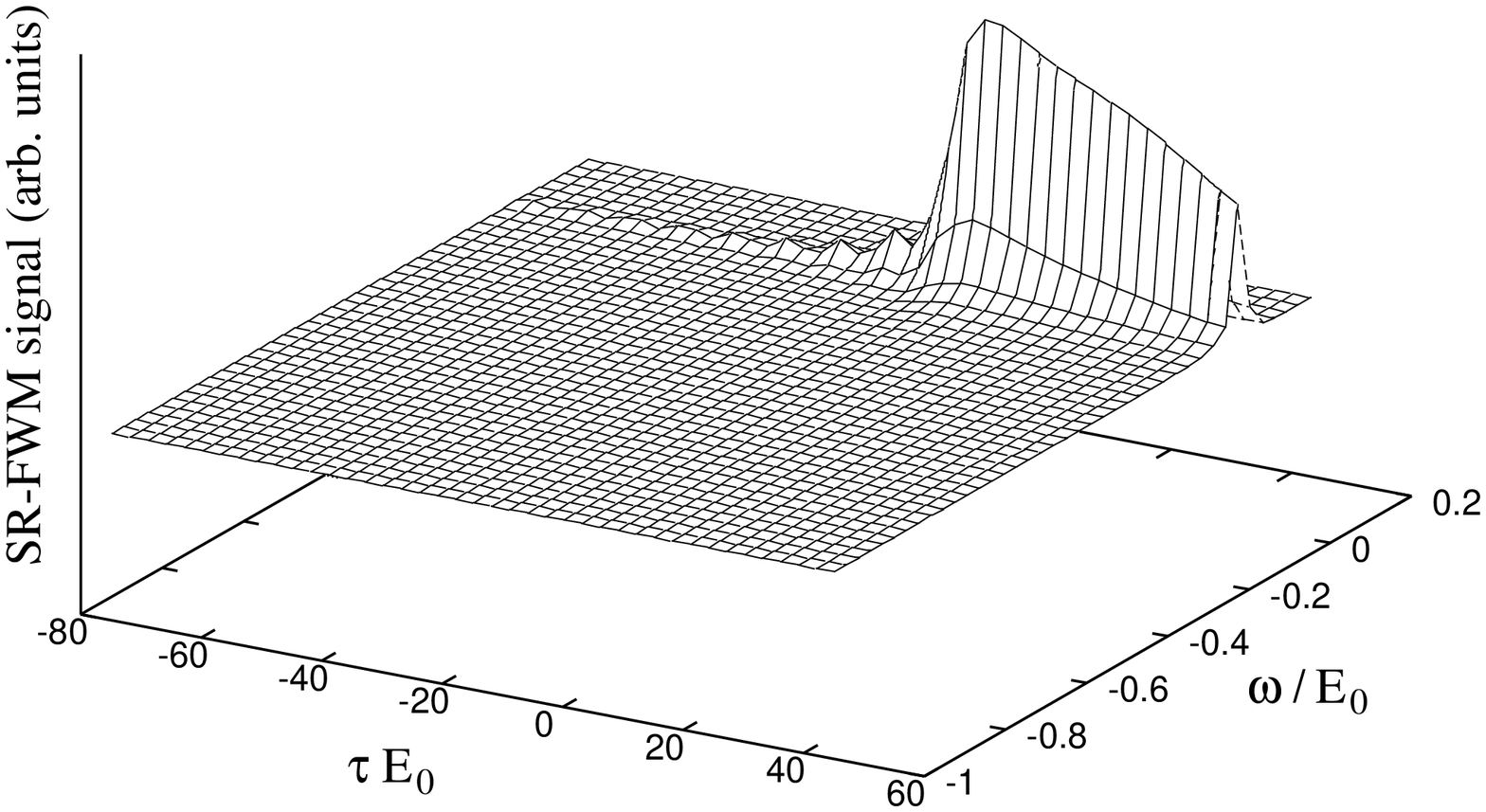}
\end{center}
\vspace{80mm}
\centerline{Fig. 6}
\clearpage
\begin{center}
\epsfxsize=6.0in
\epsffile{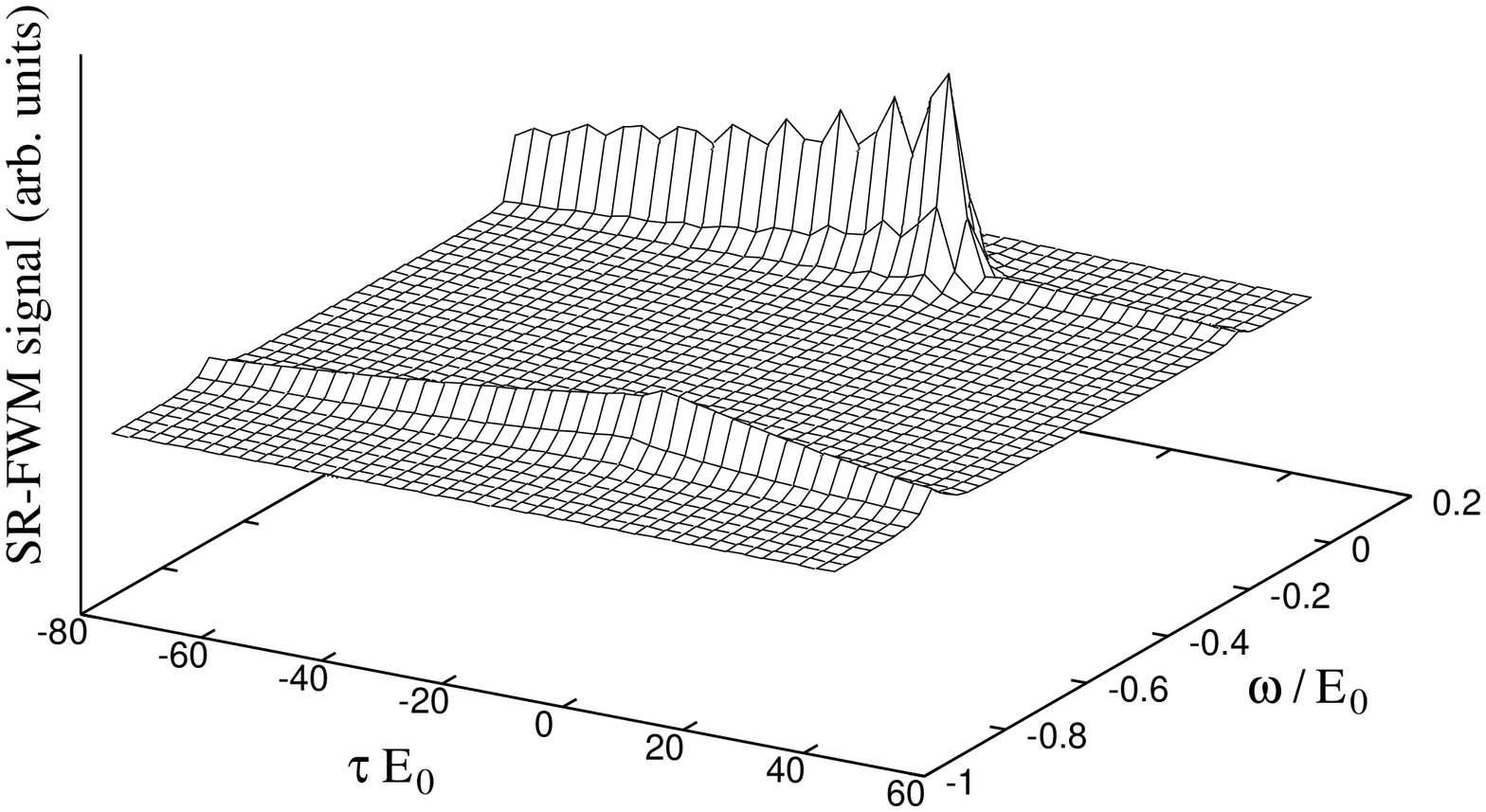}
\end{center}
\vspace{80mm}
\centerline{Fig. 7}
\clearpage
\begin{center}
\epsfxsize=5.0in
\epsffile{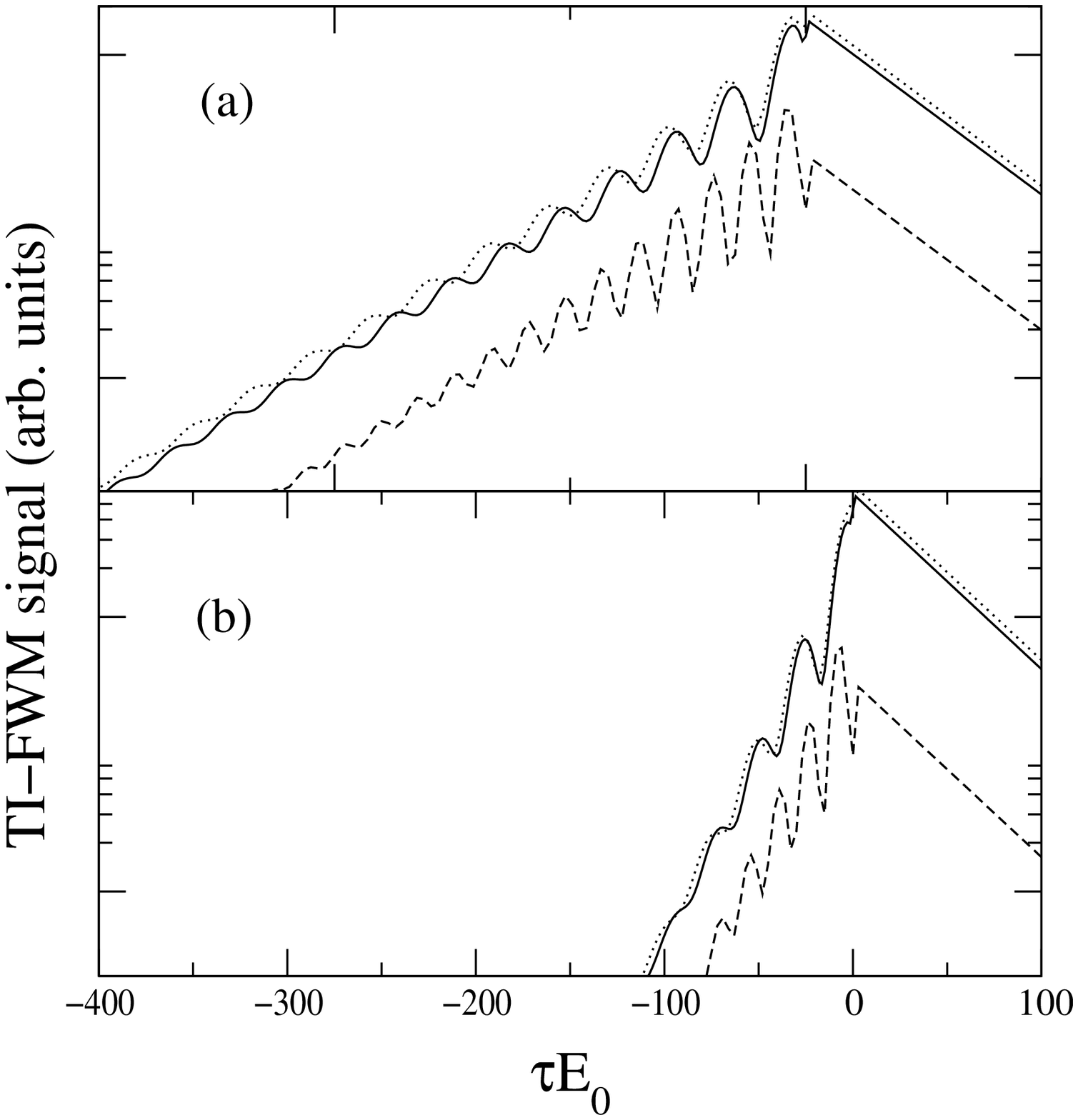}
\end{center}
\vspace{40mm}
\centerline{Fig. 8}
\clearpage
\begin{center}
\epsfxsize=5.0in
\epsffile{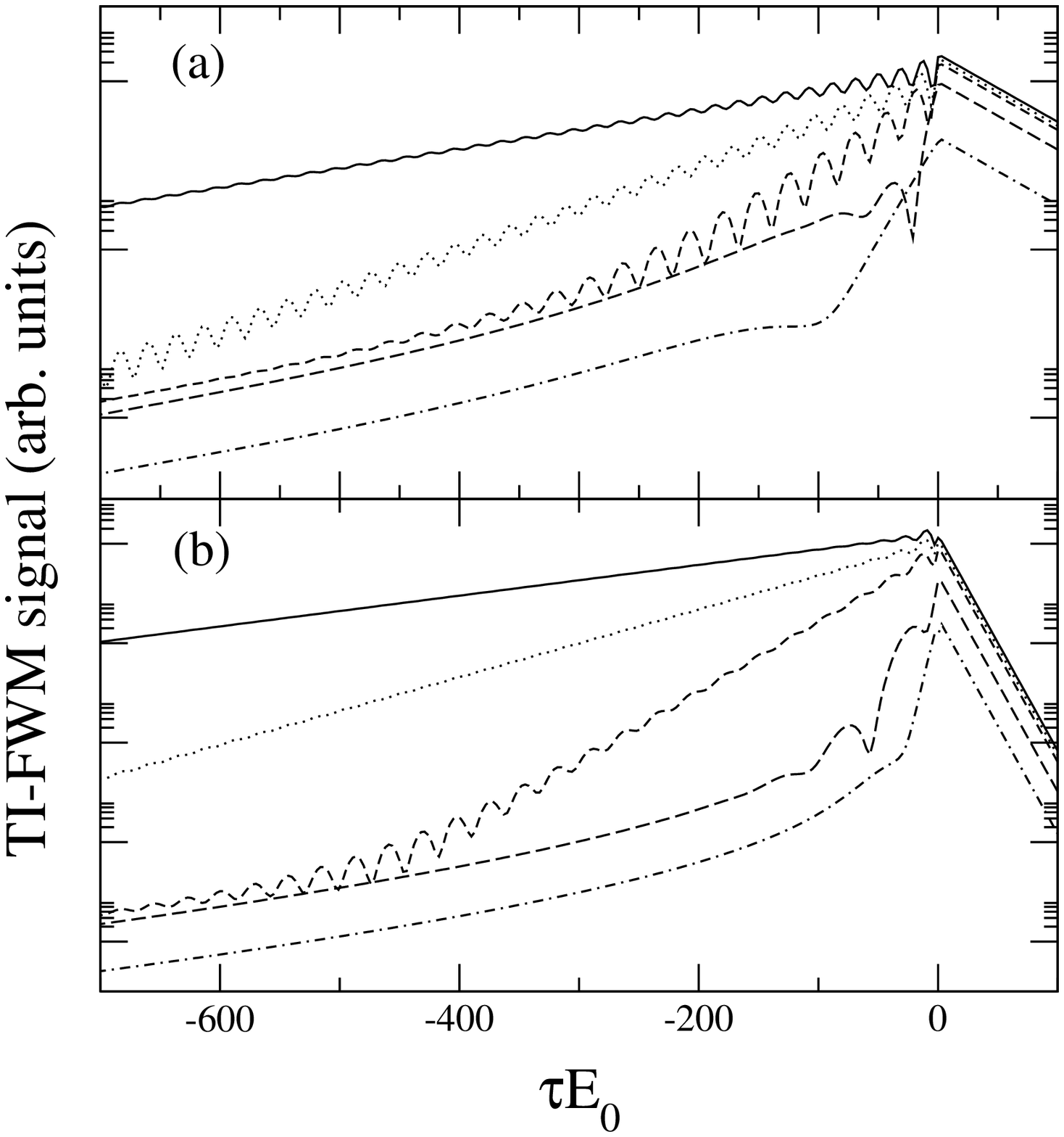}
\end{center}
\vspace{40mm}
\centerline{Fig. 9}
\clearpage
\begin{center}
\epsfxsize=5.0in
\epsffile{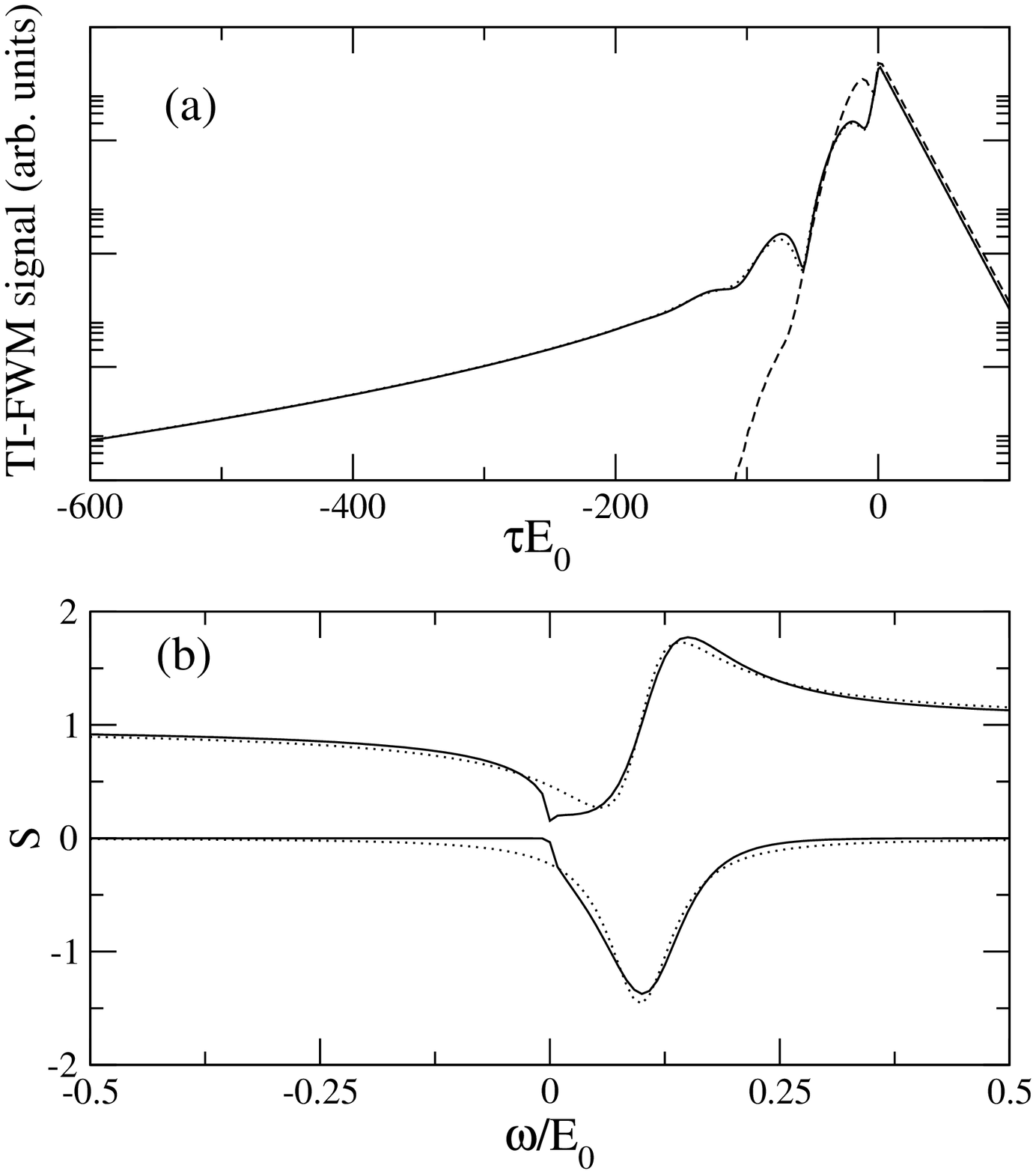}
\end{center}
\vspace{40mm}
\centerline{Fig. 10}
\clearpage
\begin{center}
\epsfxsize=5.0in
\epsffile{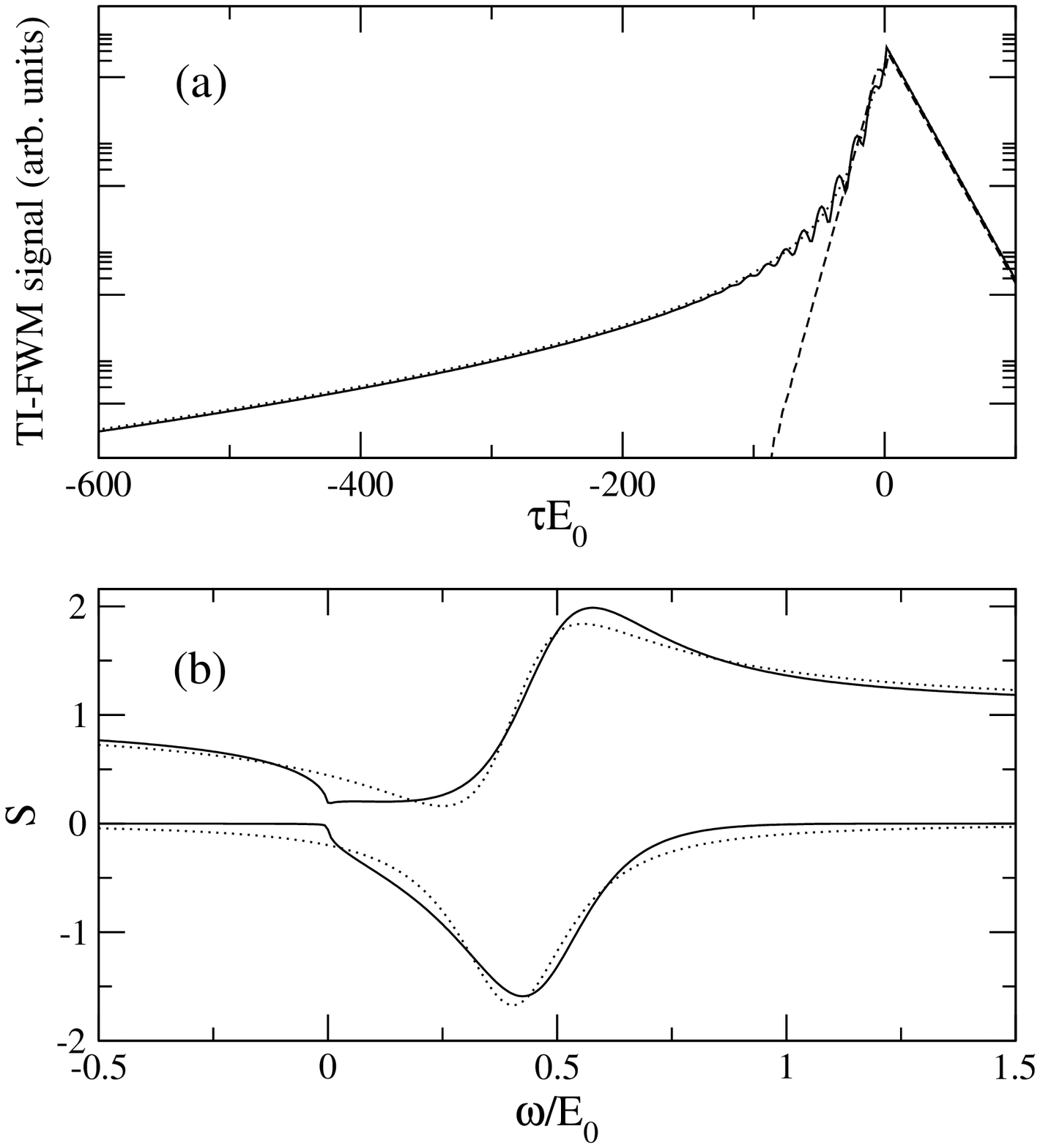}
\end{center}
\vspace{40mm}
\centerline{Fig. 11}

\end{document}